\documentclass[twocolumn, pre]{revtex4}

\usepackage{amssymb}
\usepackage{amsmath}
\usepackage{graphics}
\usepackage{graphicx}
\usepackage{mathrsfs}
\usepackage{verbatim}

\newcommand{\be}{\begin{equation}}
\newcommand{\ee}{\end{equation}}
\newcommand{\Ang}{\textrm{ \AA}}

\begin{document}

\title{Non-mean-field theory of anomalously large double-layer capacitance}

\date{\today}
\author{M. S. Loth}
\author{Brian Skinner}
\author{B. I. Shklovskii}
\affiliation{Fine Theoretical Physics Institute, University of Minnesota, Minneapolis, Minnesota 55455}

\begin{abstract}

Mean-field theories claim that the capacitance of the double-layer formed at a metal/ionic conductor interface cannot be larger than that of the Helmholtz capacitor, whose width is equal to the radius of an ion.  However, in some experiments the apparent width of the double-layer capacitor is substantially smaller. We propose an alternate, non-mean-field theory of the ionic double-layer to explain such large capacitance values.  Our theory allows for the binding of discrete ions to their image charges in the metal, which results in the formation of interface dipoles. We focus primarily on the case where only small cations are mobile and other ions form an oppositely-charged background.  In this case, at small temperature and zero applied voltage dipoles form a correlated liquid on both contacts. We show that at small voltages the capacitance of the double-layer is determined by the transfer of dipoles from one electrode to the other and is therefore limited only by the weak dipole-dipole repulsion between bound ions, so that the capacitance is very large. At large voltages the depletion of bound ions from one of the capacitor electrodes triggers a collapse of the capacitance to the much smaller mean-field value, as seen in experimental data.  We test our analytical predictions with a Monte Carlo simulation and find good agreement. We further argue that our ``one-component plasma" model should work well for strongly asymmetric ion liquids. We believe that this work also suggests an improved theory of pseudo-capacitance.

\end{abstract} \maketitle

\section{Introduction}

The rising demand for compact forms of energy storage with high
power output has resulted in increased interest in electrochemical
capacitors (ECs) \cite{Abruna, Conway}. An EC is a pair of metal
electrodes separated by an ionic conductor, such as an aqueous
solution of ions, an ionic liquid \cite{Galinski}, a super-ionic
crystal \cite{Agrawal}, or an ion-conducting glass \cite{Angell,
Mariappan}.  ECs with extremely high area per unit volume
(``supercapacitors") already have a large number of applications. In this paper our focus is not on large surface area, but on the deeper physical question of a maximum possible capacitance per unit area.

In a conventional double-plate capacitor, where metal electrodes
of area $S$ are separated by an insulator of width $d$ and
dielectric constant $\varepsilon$, the capacitance $C = \varepsilon S/4 \pi
d$ (in Gaussian units).  In an EC, the intervening medium is
actually a conductor with finite conductivity $\sigma$, but with
blocking of both ionic and electronic current at the electrode
interface.  The relation $C = \varepsilon S/ 4 \pi d$ is therefore only valid at sufficiently high frequencies $\omega \gg 4 \pi
\sigma / \varepsilon$, where the bulk of the ionic medium behaves as an
insulator.  We concern ourself with the opposite limit $\omega \ll 4 \pi \sigma / \varepsilon$, where polarization of the ionic medium eliminates electric field in the bulk and the capacitance of the EC is determined by the formation of thin electrostatic double-layers (EDLs) at both electrodes.

How large can the capacitance be for these double-layers?  The commonly-accepted expression for the maximum possible capacitance of an EDL goes back to Helmholtz \cite{Helmholtz1853}, who assumed that the charge of the metal surface is compensated by a layer of counterions with diameter $a$ residing on the surface of metal. The resulting ``Helmholtz
capacitance" is given by
\be 
C_H = \varepsilon S/2 \pi a. \label{eq:Helmholtz} 
\ee
For a double-plate capacitor, where the EDLs formed at both
electrodes can be thought of as two equal capacitances connected in
series, the maximum capacitance is $C_H/2 = \varepsilon S/4 \pi a$. For
$a = 2\textrm{\AA}$ and $\varepsilon = 5$, as we use below, $C_H/2S = 22$ $\mu$F/cm$^2$.

A recent experiment \cite{Mariappan}, however, has reported much larger values of the EDL capacitance in phosphosilicate glasses placed between platinum electrodes (see Fig$.$ \ref{fig:compare_to_glass}).  Capacitance per unit area as large as $400\hspace{1mm} \mu \textrm{F/cm}^2$  was measured, corresponding to an effective capacitor thickness $d^*=\varepsilon S/ 4 \pi C$ in the range $0.2$ -- $0.7$ \AA, much smaller than any ion radius. The glass was held at a temperature of 573 K, at which only the smallest ions, Na$^+$ with diameter $a = 2\textrm{\AA}$, are mobile.  The dielectric constant of the glass $\varepsilon$ is between 2 and 10.

Current theories of EDL capacitors, based on the mean-field approach, fail to explain such large capacitance.  The most widely-used theory is that of Gouy, Chapman, and Stern (GCS) \cite{Gouy1910,Chapman1913, Stern1924}, which extends the Helmholz capacitor concept to allow for the thermal motion of counterions.  In this approach, neutralizing ionic charge is imagined as a stack of thin uniform layers placed parallel to the charged electrode, with the charge density of each layer dictated by the Poisson-Boltzmann equation. When excluded volume of ions is
properly taken into account \cite{Kornyshev2007, Freise1952,
Borukhov, Kilic2007, Nguyen, Oldham2008} such theories lead only
to a larger effective capacitor thickness, and therefore a smaller
capacitance than the Helmholtz value.

In this paper we propose an alternate theory to explain the large
differential capacitance of the EDL. We abandon the mean-field
approximation and deal instead with discrete ion charges, which
interact strongly with the metal surface in a way that is not
captured by the mean-field approximation.  For simplicity, this paper focuses primarily on the case where only the cations are mobile, and therefore may form the EDL, while anions comprise a fixed background of negative neutralizing charge.  This ``one component plasma" (OCP) model is used here to describe the ion-conducting glasses examined in Ref$.$ \cite{Mariappan}.  We also suggest a number of other systems to which it can be applied.

One other application of the OCP model is to super-ionic crystals, where only the smallest positive ionic species (such as Na$^+$ or Li$^+$) is mobile. Less obvious is the application of the OCP model to ionic liquids, where both positive and negative ions are mobile. Nevertheless, we show below that for ionic liquids composed of monovalent cations and much larger, rigid anions, one can think about space-filling anions as a weakly-compressible negative background on which the small cations rearrange to form the EDL. If the cations are multivalent, then monovalent anions can again be treated as a negative background even if they are of the same size as the cations.  These applications will be explained in greater detail in a later section.

We begin our theory of the OCP model by noting that a cation adjacent to a metal electrode produces electronic polarization of the metal surface, and the cation experiences an attraction to the resulting image charge.  For ions of small radius, the image attraction is significantly larger than the thermal energy $k_BT$, so that ions form stable, compact ion-image dipoles at the metal surface. Cations may also experience some chemically-specific attraction to the metal electrodes, which enhances the effects of image attraction. Repulsion between adjacent dipoles results in the formation of a strongly-correlated liquid of dipoles along the surface of both electrodes.

\begin{figure}[htb]
\centering
\includegraphics[width=0.45 \textwidth]{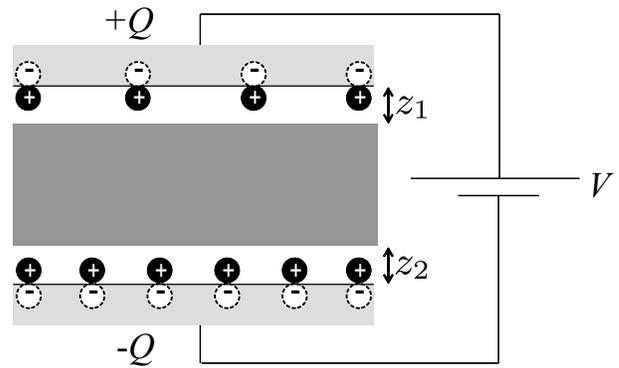}
\caption{A capacitor consisting of parallel metal plates (lightly shaded)
bounding an ionic conductor with mobile positive ions.  The neutral region of the ionic conductor is heavily shaded, while negative depletion regions ($z_1$ and $z_2$) are left white.  The relative size of the ions has been greatly exaggerated.} \label{fig:schematic}
\end{figure}

The adsorbed cations leave behind a region of negative background, or a depletion layer.  Thus, each border of the sample is spontaneously polarized in the direction perpendicular to it (see Fig$.$ \ref{fig:schematic}).  When a voltage $V$ is applied between the electrodes, these opposite-facing, spontaneous polarizations are easily rearranged in the direction of the external electric field, leading to the rapid build-up of an electronic charge $\pm Q$ on the metal surfaces.

As the charge $\pm Q$ is added to the electrodes, cations are driven to detach from the positive electrode and to bind to the negative one. Below, we demonstrate that the
resulting capacitance $dQ/dV$ of the two EDLs can be significantly larger than the Helmholtz value $C_H/2$, since the dipole-dipole repulsion
that resists ion transfer is relatively small.  In other words, the effective capacitor thickness $d^*$ can be much smaller than the ion radius. Below we derive an expression for $d^*$ that is reasonably close to experimental values.  Our theory also explains another peculiarity of the experiment \cite{Mariappan}, namely the sharp drop in capacitance at a certain critical voltage (as in Fig$.$ \ref{fig:compare_to_glass}). We show that, indeed, the capacitance should collapse to a much smaller value at a particular nonzero voltage. Contrary to the mean-field theories developed in Refs$.$ \cite{Kornyshev2007, Freise1952, Borukhov, Kilic2007, Nguyen, Oldham2008}, this drop-off (shoulder) in capacitance is not driven by excluded volume effects among bound ions. Rather, it occurs far below the complete filling of an ionic layer at either electrode \cite{Baldelli2008}, when the voltage difference induces the positively-charged electrode to lose all of its bound ions.

The following section develops the OCP model of capacitance for a parallel double-plate capacitor at low temperatures, where all
entropic effects can be ignored. Section III extends this
low-temperature theory to the case of an asymmetric double-plate
capacitor, where the anode and cathode have different areas. In
the limit of one area going to infinity, this includes the description of a single EDL capacitance. Section IV examines the temperature-dependence of the capacitance, and section V presents the results of a simple Monte Carlo simulation and compares them with our analytic theory. Section VI examines the application of the OCP model to ionic liquids, and section VII explores its relevance for electrolyte solutions in water.  In Section VIII we summarize our main conclusions and we argue that our paper suggests an improved theory of ``pseudo-capacitance".

A short version of this paper was published in Ref.\ \cite{us-PRL}.

\section{Low-temperature theory of a symmetric parallel-plate capacitor}  \label{sec:lowT}

We consider the case of a parallel-plate capacitor, where an ion-conducting medium is bounded on two sides by metal plates which are maintained at a relative potential difference $V$ by some voltage source.  The intervening medium is modeled as a fixed negative background with charge density $-e N$, upon which resides a neutralizing concentration of mobile positive ions with charge $+e$ and bulk density $N$.  We model the conducting ions as hard spheres with diameter $a$, and this paper generally assumes that $Na^3 \ll 1$.  When an ion is up against the metal surface, it experiences an attractive potential energy of approximately 
\be 
u_{im} = -\frac{e^2}{2 \varepsilon a} \label{eq:Uimage}. 
\ee 
For $T = 573$ K, $\varepsilon = 5$, and $a = 2$ \r{A}, as estimated for the experiments of Ref$.$ \cite{Mariappan}, we get $|u_{im}|/k_BT \approx 15$, so that such ions are bound strongly to the surface.  In fact, the attraction of ions to the metal surface may be stronger than $u_{im}$, either because of local behavior of the dielectric constant or because of some specific chemical affinity that ions may have for the metal surface.  In general, we will write the strength of ion attraction to the metal surface as $\gamma u_{im}$, where $\gamma$ is a positive constant of order one.  Unless otherwise stated, all numerical estimates will use $\gamma = 1$. 

At a given voltage $V$, some area densities $n_1$ and $n_2$ of ions bind to the anode and cathode, respectively.  The attachment of these positive ions to the metal surface results in the formation of regions with negative net charge $q_1$ and $q_2$ near the anode and cathode, respectively. Each of these charges exactly cancels the net charge of the adjoining plate and its bound ions, so that there is no electric field in the bulk of the ionic conductor.  This implies $q_1 = -en_{1}S - Q$ and $q_2 = - en_{2}S +Q$, where $Q$ is the electronic charge that moves through the voltage source. In other words, the dipoles at each metal-glass interface are effectively embraced by a capacitor composed of the charge $q$ and its positive image $-q$ in the metal. As in every plane capacitor, the charges $q_1$ and $q_2$ are uniformly distributed along the plane.

Since positive ions gain a large energy $|u_{im}|$ by adsorbing to the metal surface, at equilibrium there must be a correspondingly large potential difference $V_{im} \equiv |u_{im}|/e$ between the metal surface and the bulk of the ionic conductor, so that the chemical potential of ions is uniform.  This large potential difference can be defined as the potential of zero charge (PZC), which has a nonzero value because of the strong spontaneous polarization of the system.  In this paper we consistently use the external electrode-electrode potential $V$, which by definition is zero when there is no charge $Q$ on the electrode surfaces, to define capacitance, rather than the potential difference between the electrode and the bulk.  

With such a large internal potential difference $V_{im} \gg k_BT/e$ at each electrode, the negative regions are strongly depleted of ions even at very small applied voltage $V$.  The charges $q_1$ and $q_2$ therefore constitute depletion layers of width $z_1$ and $z_2$ which form at the anode and cathode, respectively; here it is assumed that $z_1, z_2 \gg a$.  These layers are devoid of mobile ions and have a charge density equal to that of the negative background, so that $q_1 = -eNz_1$ and $q_2 = -eNz_2$.  Thus
\begin{eqnarray}
e N z_1 - e n_1 & = & Q/S \\
e N z_2 - e n_2 & = & - Q/S,
\end{eqnarray}
where $Q$ is the electronic charge that moves through the voltage source of the capacitor.  The electrostatic energy associated with the formation of the depletion layers and their corresponding positive image charge in the metal can be estimated as
\be
U_{dep} = \frac{2 \pi e^2 S} {3 \varepsilon N} \left[ \left(n_1 + \frac{Q}{Se}\right)^3 + \left(n_2 - \frac{Q}{Se}\right)^3 \right] . \label{eq:Udep}
\ee

\begin{figure}[htb]
\centering
\includegraphics[width=0.45 \textwidth]{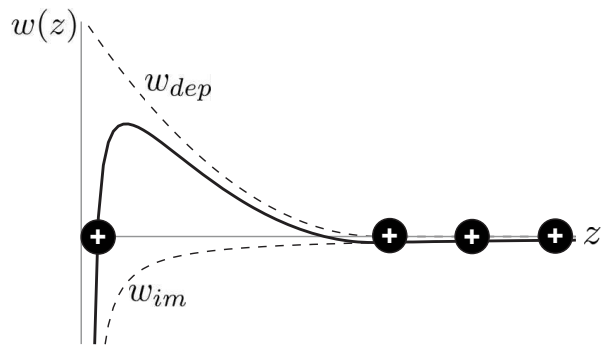}
\caption{Schematic depiction of the potential energy $w(z)$ of an ion in the vicinity of the metal plate.  The image attraction energy $w_{im}$ and the energy associated with moving through the depletion layer $w_{dep}$ are shown as a function of distance $z$ from the metal surface, along with the total $w = w_{im} + w_{dep}$.  At equilibrium, ions bound to the surface have the same potential energy as ions in the bulk.} \label{fig:schematic-potential}
\end{figure}

Fig.\ \ref{fig:schematic-potential} gives a schematic depiction of the potential energy $w(z)$ of an ion near the anode as a function of its distance $z$ from the metal surface, relative to a position in the bulk of the ionic conductor.  There are two main contributions to $w(z)$: the attraction energy $w_{im}(z) = -e^2/4 \varepsilon z$ of the ion to its image charge and the electrostatic energy $w_{dep}(z) = 2 \pi e^2 N (z_1 - z)^2/\varepsilon$ associated with moving the ion from the bulk into the depletion region.  At equilibrium, these two contributions satisfy $w_{im}(a/2) = u_{im} = -w_{dep}(a/2)$, so that bound ions at the surface have the same energy as ions in the bulk and the chemical potential is uniform.

Eq$.$ (\ref{eq:Udep}) assumes that positive bound ions are effectively neutralized by their negative image charges, so that these ions can be ignored in calculating the electrostatic self-energy of the depletion region and its image charge.  One can consider, however, the small correction to Eq$.$ (\ref{eq:Udep}) resulting from the interaction between bound ion-image pairs and the negative background.  In the limit $z_1, z_2 \gg a$, this correction can be viewed as the potential energy of a collection of ion-image dipoles, each with dipole moment $|\vec{p}| = e a$, aligned with an external field $\vec{E}$ produced by the negative depletion layer and its positive image charge.  If the number of dipoles on a given metal surface is $nS$, then the total interaction energy of these dipoles with the external field is $-nS \vec{p} \cdot \vec{E}$, so that to the energy associated with forming the depletion layers there should be added a term
\be 
U_{dE} = -4 \pi u_{im} S \left[ n_1 \left(n_1 + \frac{Q}{e S}\right) + n_2 \left(n_2 - \frac{Q}{e S} \right) \right] a^2.
\ee

In addition to the ions' interaction with the background, there is a positive energy associated with repulsion between bound ions.  When the density of bound ions is low enough that $n a^2 \ll 1$, ions repel each other by a dipole-dipole interaction: the potential created by a bound ion and its image charge repels an adjacent ion.  In
this limit the repulsive interaction between two adjacent bound ions is 
\be 
u_{dd}(n) \simeq e^2 a^2 n^{3/2} / 2 \varepsilon \label{eq:dipole}. 
\ee 
This repulsion results in the formation of a strongly-correlated liquid of ions on the electrode surface, reminiscent of a two-dimensional Wigner crystal \cite{Bockris}.  The total dipole energy at a given plate is $\alpha n S u_{dd}(n)$, where $\alpha$ is a numerical coefficient which depends on the structure of the lattice of dipole positions.  Thus, the total dipole interaction energy can be written as
\be 
U_{dd} = \alpha S [ n_1 u_{dd}(n_1) + n_2 u_{dd}(n_2)]. \label{eq:totaldipole}
\ee
For the minimum energy triangular lattice, $\alpha \approx 4.4$.  Further calculations will use this value.

We now assemble a full description of the total energy $U$ associated with the bound charge densities $n_1$ and $n_2$, taking as our reference the case where $n_1 = n_2 = 0$: 
\be 
U = S(n_1 + n_2) \gamma u_{im} + U_{dep} + U_{dE} +  U_{dd} - Q V. \label{eq:Utot} 
\ee
Here, $-Q V$ represents the work done by the voltage source. 

It should be noted that this expression neglects another potential contribution to the total energy: that of the finite chemical potential of ions in the bulk.  At zero temperature, for example, ions in the bulk are crystallized into a three-dimensional Wigner crystal with finite
self-energy $u_{WC} \approx -1.4 e^2 N^{1/3}/\varepsilon$ per ion \cite{Mahan}.  The energy $U$ is defined relative to the state where no ions have left the bulk, so the attachment of $(n_1 + n_2)S$ ions to the electrodes should involve an additional energy cost of $-(n_1 + n_2) S u_{WC}$.  At small $N$, this energy provides only a small correction to the binding energy $u_{im}$ and can be effectively absorbed into the coefficient $\gamma$.
 
At equilibrium, the values of $Q(V)$, $n_1(V)$, and $n_2(V)$ are those which minimize $U$.  They can therefore be found by the conditions $\partial U/\partial Q = \partial U/\partial n_1 = \partial U/\partial n_2 = 0$, the latter two of which produce
\be
\frac{5 \alpha}{2} n_1^{3/2}a^3 + \frac{4 \pi}{N a^3} \left(n_1 + \frac{Q}{e S} \right)^2 a^4 = \gamma + 4 \pi \left(2 n_1 +  \frac{Q}{e S}\right) a^2, \label{eq:dUdn1} 
\ee
\be
\frac{5 \alpha}{2} n_2^{3/2}a^3 + \frac{4 \pi}{N a^3} \left(n_2 - \frac{Q}{e S} \right)^2 a^4 = \gamma + 4 \pi \left(2 n_2 -  \frac{Q}{e S} \right) a^2. \label{eq:dUdn2}
\ee
If ions are relatively sparse in the bulk, so that $Na^3 \ll 1$, then the first and final terms of both equations can be neglected.  Setting $Q = 0$ suggests that at zero voltage, when the net charge of the capacitor is zero, there is still a finite concentration $n_0$ of ions bound to each plate given by
\be
n_0 a^2 \simeq \sqrt{\frac{\gamma N a^3} {4 \pi} } \ll 1.  \label{eq:n0}
\ee
Since the ``filling factor" $n a^2$ on each metal surface is small, bound ions are sufficiently distant that our approximation of a dipole interaction between them is justified.  This condition also verifies our assumption that the first and final terms of Eqs$.$ (\ref{eq:dUdn1}) and (\ref{eq:dUdn2}) are much smaller than unity.

As the voltage is increased from zero, ions are driven away from the anode and attracted to the cathode, so that $n_1$ decreases and $n_2$ increases.  Eqns$.$ (\ref{eq:dUdn1}) and (\ref{eq:dUdn2}) imply that $n_1$ and $n_2$ depend linearly on $Q$:
\begin{eqnarray}
n_0 - n_1 \simeq Q/eS, \label{eq:n1Q} \\
n_2 - n_0 \simeq Q/eS. \label{eq:n2Q}
\end{eqnarray}
Subtraction of Eqs$.$ (\ref{eq:n1Q}) and (\ref{eq:n2Q}) suggests that the total number of adsorbed ions $n_1 + n_2 \simeq 2n_0$ per unit area remains almost constant with increasing voltage.  This result is a direct consequence of the large difference between the depletion layer energy $U_{dep}$ and the dipole-dipole energy $\alpha S n u_{dd}(n)$.  Indeed, by comparing Eqs$.$ (\ref{eq:Udep}) and (\ref{eq:totaldipole}), we see that the condition $N a^3 \ll 1$ implies that the energy cost associated with increasing the total number of bound ions, and thereby causing the depletion layers to swell, is much larger than the dipole-dipole interaction energy.  As a consequence, the total number of bound ions remains nearly constant with voltage.  The electronic charge $Q \simeq e S (n_2 - n_1)/2$ that passes through the voltage source can therefore be thought of as the corresponding movement of image charges from one plate to another.

When enough charge has moved through the voltage source that $Q = e S n_0$, the anode becomes completely depleted of bound ions, so that $n_1 = 0$ and $n_2 \simeq 2 n_0$.  This corresponds to a particular voltage $V_c$.  At voltages $V > V_c$, the number of bound ions on the non-depleted electrode may still increase, but only through the costly widening of the depletion layer.

A relation between charge $Q$ and voltage $V$ can be derived by substituting Eqs$.$ (\ref{eq:n1Q}) and (\ref{eq:n2Q}) into Eq$.$ (\ref{eq:Utot}), so that the total energy $U(Q)$ is written as a function of the charge only.  A subsequent application of the condition $\partial U/\partial Q = 0$ gives
\be 
V \simeq \frac{5 \alpha}{2} \left[ \left(n_0 + \frac{Q}{e S}\right)^{3/2} - \left(n_0 - \frac{Q}{e S}\right)^{3/2} \right] a^3 V_{im}.
\label{eq:VQ}
\ee
By taking the derivative of this expression with respect to $V$, we can derive the capacitance $C = dQ/dV$.  This gives
\begin{eqnarray}
C & = & \frac{8}{15 \alpha} \left[ \left(\frac{\varepsilon S}{a \sqrt{n_0a^2 - Qa^2/eS}}\right)^{-1} + \nonumber \right. \\
&  &  \left. \left(\frac{\varepsilon S}{a \sqrt{n_0 a^2 + Qa^2/eS}} \right)^{-1} \right]^{-1}. \label{eq:CQ}
\end{eqnarray}
In this expression the terms inside the parentheses represent the capacitance of the anode and cathode, respectively, which add like capacitors in series.  At zero voltage, $Q = 0$, so that the capacitance becomes
\be 
C(0) \simeq \frac{4}{15 \alpha} \left( \frac{4 \pi }{\gamma N a^3} \right)^{1/4} \frac{\varepsilon S}{a} = \frac{8 \pi}{15 \alpha} \left( \frac{4 \pi}{\gamma N a^3} \right)^{1/4} C_H \label{eq:C0}.
\ee
For $Na^3 \ll 1$ one gets $C(0) \gg C_H/2$ because at small voltages charging of the capacitor is limited only by the dipole-dipole repulsion energy. Since $n_0 a^2 \ll 1$, the dipole-dipole interaction is weak, so that the resulting capacitance can be large.

At higher voltages, the capacitance of the EDL near the positive plate increases strongly as this plate plate becomes depleted of ions and the corresponding dipole repulsion energy goes to zero.  Thus the contribution of the positive plate to the total capacitance vanishes when $Q = e n_0 S$.  The corresponding voltage $V_c$ can be found by substituting $Q = e n_0 S$ into Eq$.$ (\ref{eq:VQ}):
\be 
V_c \simeq \frac{5 \alpha}{2} \left( \frac{\gamma N a^3}{\pi} \right)^{3/4} V_{im}. \label{eq:Vc} 
\ee
Immediately prior to $V = V_c$, the capacitance achieves its maximum value 
\be 
C_{max} \simeq \frac{8}{15 \alpha} \left( \frac{\pi}{\gamma N a^3 } \right)^{1/4} \frac{\varepsilon S}{a} = \sqrt2 C(0) . 
\label{eq:Cmax}
\ee 

The effective thickness $d^*_{min}$ corresponding to $C_{max}$ is
\be 
d^*_{min} \simeq  \frac{15 \alpha}{32 \pi} \left( \frac{\gamma N a^3}{\pi} \right)^{1/4} a \approx 0.49 \left(Na^3\right)^{1/4} a . 
\ee 
Thus we arrive at a remarkable prediction: the effective capacitor thickness can be much smaller than the ion diameter $a$. As an example, an ion-conducting medium with mobile sodium atoms of diameter $a = 2$ \AA ~ and density $N a^3 = 0.01$ can be used to make a capacitor with capacitance nearly six times larger than $C_H/2$. As mentioned before, this surprisingly high capacitance is a result of the weak dipole-dipole interaction between bound ions that comprise the double-layer.  Indeed, near the capacitance maximum the filling factor on the negative plate $n_2 a^2 \simeq 2 n_0 a^2 \ll 1$, so that it is incorrect to think of the EDL as a series of uniformly charged layers. This difference represents an important change of paradigm, from a mean-field capacitor to a capacitor composed of discrete, correlated dipoles.

At $V > V_c$, ions can no longer simply be transferred from anode to cathode, and the capacitance collapses to a much smaller value.  The value of this ``depleted capacitance" can be found through optimization of the total energy $U$ under the condition $n_1 = 0$, which yields 
\be 
C_{dep}(V) \simeq  \frac{\varepsilon S}{a} \sqrt{\frac{N a^3 }{4 \pi (V/V_{im} + \gamma)} } . \label{eq:Cdep}
\ee
This expression neglects the weak dipole-dipole interaction at the non-depleted negative plate.  At $V/V_{im} \gg \gamma$ the capacitance is dominated by the depletion layer next to the positive electrode, and therefore it approaches the standard value for capacitance of a depletion layer.

\begin{figure}[htb]
\centering
\includegraphics[width=0.45 \textwidth]{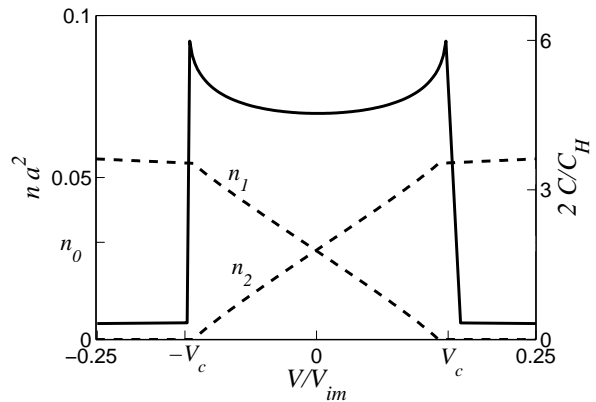}
\caption{The area densities of bound ions (dashed lines, left vertical axis) and the capacitance (solid line, right vertical axis) as a function of applied voltage for $Na^3 = 0.01$, as given by a numerical minimization of the total energy in Eq$.$ (\ref{eq:Utot})} \label{fig:c-n}
\end{figure}

Fig$.$ \ref{fig:c-n} shows the capacitance and the density of bound ions as a function of voltage, as calculated by a numerical minimization of the total energy in Eq$.$ (\ref{eq:Utot}).  We have used $Na^3 = 0.1$, following the estimate of Ref$.$ \cite{Mariappan}.  All results are within 10\% of the approximate analytic expressions in Eqs$.$ (\ref{eq:n0}) -- (\ref{eq:n2Q}), (\ref{eq:CQ}) -- (\ref{eq:Cdep}).

\section{Low-temperature theory of an asymmetric Capacitor}

We may also consider the case of an asymmetric capacitor, where the two electrodes have unequal areas.  This may be the case, for example, in a coaxial cylindrical capacitor where the inner electrode has a smaller radius than the outer electrode.  For the sake of argument, we assume that at positive voltages the smaller electrode (area $S_1$) is the anode and the larger electrode (area $S_2$) is the cathode.  The corresponding Helmholtz capacitance for the double layer at each electrode is $C_{H,1} = \varepsilon S_1/2 \pi a$ and $C_{H,2} = \varepsilon S_2/2 \pi a$ for the small and large electrodes, respectively.  These two capacitances add in series, so that by the GCS theory the maximum possible capacitance is
\be 
\left(\frac{1}{C_{H,1}} + \frac{1}{C_{H,2}} \right)^{-1} = 
\frac{\varepsilon}{2 \pi a} \left(\frac{1}{S_1} + \frac{1}{S_2} \right)^{-1}. \label{eq:CHseries}
\ee
When $S_1 = S_2 = S$, this result reduces to $C_H/2$, as in the previous section.  In the limit $S_2 \gg S_1$, Eq$.$ (\ref{eq:CHseries}) approaches $C_{H,1}$, so that the total capacitance is limited by the smaller area.

Let us now see how our theory of the previous section can be generalized to the asymmetric case.  The conditions of neutrality at anode and cathode become
\begin{eqnarray}
e N z_1 - e n_1 = Q/S_1 \\
e N z_2 - e n_2 = - Q/S_2,
\end{eqnarray}
so that the electrostatic energy of the depletion layers can be written as
\be
U_{dep} = \frac{2 \pi e^2} {3 \varepsilon N} \left[ S_1 \left(n_1 + \frac{Q}{S_1e}\right)^3 + S_2 \left(n_2 - \frac{Q}{S_2e}\right)^3 \right] .
\ee
Similarly, the interaction between dipoles and the depletion layer becomes
\be 
U_{dE} = -4 \pi u_{im} \left[ n_1 S_1 \left(n_1 + \frac{Q}{e S_1} \right) + n_2 S_2 \left( n_2 - \frac{Q}{e S_2} \right) \right] a^2.
\ee
and the total dipole-dipole interaction energy is
\be 
U_{dd} = \alpha \left[S_1 n_1 u_{dd}(n_1) + S_2 n_2 u_{dd}(n_2) \right].
\ee
The total energy is therefore
\be
U = (S_1 n_1 + S_2 n_2) \gamma u_{im} + U_{dep} + U_{dE} + U_{dd} - Q V. \label{eq:Utot2} 
\ee
As before, the values of $Q(V)$, $n_1(V)$, and $n_2(V)$, are those which minimize $U$.

At zero voltage, the conditions $\partial U/\partial n_1 = \partial U/\partial n_2 = 0$ suggest $n_1 = n_2 = n_0$, where $n_0$ is the same as in the symmetric case [Eq$.$ (\ref{eq:n0})].  At finite $V$ they imply
\begin{eqnarray}
n_0 - n_1 & = & Q/eS_1 \label{eq:n1QS}\\
n_2 - n_0 & = & Q/eS_2, \label{eq:n2QS}
\end{eqnarray}
so that, as in the symmetric case, the charge $Q$ can be thought of as the movement of image charges from one electrode to another.  Here we have again assumed that $Na^3 \ll 1$.

Using Eqs$.$ (\ref{eq:n1QS}) and (\ref{eq:n2QS}), we can rewrite the total energy as a function of $Q$ only.  By optimizing $U(Q)$ with respect to $Q$ we obtain an expression $V(Q)$ from which we can define the capacitance.  This procedure yields
\be 
V \simeq \frac{5 \alpha}{2} \left[ \left(n_0 + \frac{Q}{e S_2}\right)^{3/2} - \left(n_0 - \frac{Q}{e S_1}\right)^{3/2} \right] a^3 V_{im}.
\label{eq:VQS}
\ee
Taking the derivative with respect to $V$, and using $C = dQ/dV$, we get
\begin{eqnarray}
C & = & \frac{8}{15 \alpha} \left[ \left(\frac{\varepsilon S_1}{a \sqrt{n_0a^2 - Qa^2/eS_1}}\right)^{-1} + \nonumber \right. \\
&  &  \left. \left(\frac{\varepsilon S_2}{a \sqrt{n_0 a^2 + Qa^2/eS_2}} \right)^{-1} \right]^{-1}. \label{eq:CQS}
\end{eqnarray}
In this expression the terms inside the parentheses represent the capacitance of the anode and cathode, respectively, which add like capacitors in series.  At zero voltage, $Q = 0$, so that the capacitance becomes
\be 
C(0) \simeq \frac{8}{15 \alpha} \left( \frac{4 \pi}{\gamma N a^3} \right)^{1/4} \frac{\varepsilon}{a} \left(\frac1{S_1} + \frac1{S_2} \right)^{-1}. \label{eq:C0S}
\ee
When $S_1 = S_2$, this expression reduces to the symmetric result of Eq$.$ (\ref{eq:C0}).  When $S_1 \ll S_2$, the capacitance is dominated by the smaller area, as expected.  In either case, $C(0)$ is again much larger than the corresponding Helmholtz values $C_H/2$ and $C_{H,1}$.

Below we consider the behavior of the capacitance as a function of voltage, examining separately the cases of positive and negative voltage applied to the small electrode.

\subsection{Positive voltage}

When the voltage is increased from zero, the capacitance increases as ions unbind from the anode and bind to the cathode.  At a certain critical voltage $V_{c,1}$ the anode becomes depleted of bound ions.  This occurs when $n_1 = 0$, or $Q = e S_1 n_0$, so that by Eq$.$ (\ref{eq:VQS})
\be 
V_{c,1} \simeq \frac{5 \alpha}{2} \left( \frac{\gamma N a^3}{4 \pi} \right)^{3/4} \left(1 + \frac{S_1}{S_2} \right)^{3/2} V_{im}.
\label{eq:Vc1}
\ee
At this point, the double-layer capacitance of the anode diverges as ions bound to the anode become sparse and the corresponding dipole-dipole interaction energy goes to zero.  The total capacitance is therefore limited only by the capacitance of the cathode.  Substituting $Q = e S_1 n_0$ into Eq$.$ (\ref{eq:CQS}) gives
\be 
C(V_{c,1}) \simeq \frac{S_2}{S_1} \sqrt{ \frac{S_1 + S_2}{S_2} } C(0).
\label{eq:CVc1}
\ee
When the cathode is much larger than the anode, $S_2/S_1 \gg 1$, the maximum capacitance $C(V_{c,1}) \simeq C(0) S_2/S_1$.  This result implies a tremendous growth in the capacitance at positive voltages $0 < V < V_{c,1}$.  For voltages just below the critical value, so that $V_{c,1} - V \ll V_{c,1}$, we can examine this growth analytically.  
At $S_2/S_1 \gg 1$ we can ignore the term $Q/eS_2$ in Eq. (\ref{eq:VQS})
and we arrive at $(n_0 a^2 - Q  a^2/eS_1)^{3/2} = (V_{c,1} - V)/ V_{im}$.
Substituting this result for the first term of the sum in Eq$.$ (\ref{eq:CQS}) (the inverse capacitance of the smaller electrode) and ignoring the second term (the inverse capacitance of the larger electrode), we find that the capacitance diverges approximately as
\be 
C(V) \simeq \frac43 \left( \frac{2}{5 \alpha} \right)^{2/3} \left( \frac{V_{im}}{V_{c,1} - V} \right)^{1/3} \frac{\varepsilon S_1}{a}, \hspace{0.5cm} (0 < V < V_{c,1}) \label{eq:Cdiv}
\ee
before being truncated by the finite value of $S_2$ as given in Eq$.$ (\ref{eq:CVc1}).  At $V = 0$, Eq$.$ (\ref{eq:Cdiv}) approximately matches the capacitance $C(0)$ from Eq$.$ (\ref{eq:C0}).

At larger voltages $V > V_{c,1}$, the anode becomes depleted of ions and the capacitance collapses to a much smaller value.  This value can be found through minimization of the total energy $U$ with respect to $Q$ under the condition $n_1 = 0$, which gives
\be 
C_{dep, 1}(V) \simeq \sqrt{\frac{Na^3}{4 \pi(V/V_{im} + \gamma )}} \frac{\varepsilon S_1}{a} , \hspace{0.5cm} (V > V_{c,1}) \label{eq:Cdep1}.
\ee 
In other words, the capacitance at large positive voltages is dominated by that of the growing depletion layer at the anode, as in Eq$.$ (\ref{eq:Cdep}) for the symmetric case.

\begin{figure}[htb]
\centering
\includegraphics[width=0.45 \textwidth]{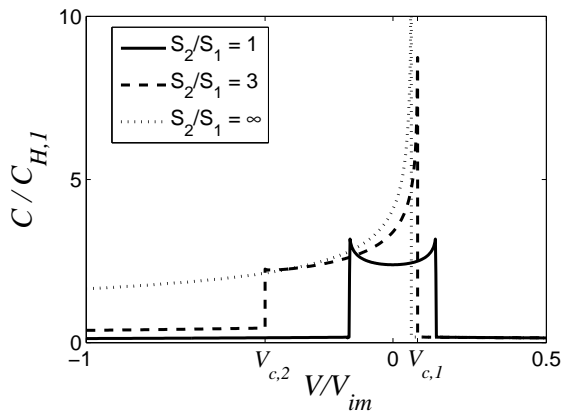}
\caption{The capacitance as a function of voltage for different ratios $S_2/S_1$, as calculated by numerical minimization of the total energy, for $Na^3 = 0.01$.  Threshold voltages $V_{c,1}$ and $V_{c,2}$ are shown for the dashed line only.}
\label{fig:area_ratios}
\end{figure}

Fig$.$ \ref{fig:area_ratios} shows the capacitance as a function of voltage for different values of $S_2/S_1$, as calculated by numerical minimization of the total energy in Eq$.$ (\ref{eq:Utot2}).  For $V > 0$, this figure illustrates the analytical results of Eqs$.$ (\ref{eq:Vc1}) -- (\ref{eq:Cdep1}).  For $V < 0$ it shows the predictions of the following subsection.  Notice that at large values of $S_2/S_1$ the divergence in the capacitance near $V = V_{c,1}$ becomes increasingly pronounced.

\subsection{Negative voltage}

So far we have been talking about the case of a small anode and large cathode.  We now consider the situation of negative voltage $V$, where the electrodes change roles and the cathode area $S_1$ is smaller than the anode area $S_2$.  This change in electrode roles suggests an asymmetry in the total capacitance with respect to the sign of the voltage.  Below we examine the capacitance at negative voltages for two limiting cases: the moderately asymmetric case $S_2/S_1 \ll 1/\sqrt{Na^3}$ and the highly asymmetric case $S_2/S_1 \gg 1/\sqrt{Na^3}$.  

\subsubsection{Moderately asymmetric capacitor}

If the two electrode areas are not too different, so that $S_2/S_1 \ll 1/\sqrt{Na^3}$, then the capacitance at negative voltages can be described using the same procedure as for positive voltage, beginning with Eqs$.$ (\ref{eq:VQS}) -- (\ref{eq:CQS}).  The voltage $V_{c,2}$ at which the larger electrode becomes depleted of bound ions can be found by setting $Q = -e S_2 n_0$ in Eq$.$ (\ref{eq:VQS}), which gives
\be 
V_{c,2} = - \left(\frac{S_2}{S_1}\right)^{3/2} V_{c,1}.
\ee
The large electrode therefore becomes depleted at larger voltages (in absolute value) than the small electrode (see the dashed line in Fig$.$ \ref{fig:area_ratios}).  The condition $S_2/S_1 \ll 1/\sqrt{Na^3}$ guarantees that $|V_{c,2}| < V_{im}$.

At larger negative voltages $V < V_{c,2}$, there is an apparent collapse in capacitance due to the depletion of bound ions at the larger electrode.  The capacitance immediately prior to this voltage 
\be 
C(V_{c,2}) \simeq \frac{S_1}{S_2} \sqrt{\frac{S_1 + S_2}{S_1} } C(0) = \left( \frac{S_1}{S_2} \right)^{3/2} C(V_{c,1}).
\ee
This value is smaller than the maximum capacitance $C(V_{c,1})$ since the capacitance at $V_{c,2}$ is limited by the capacitance of the smaller electrode while at $V_{c,1}$ it is limited by the larger electrode.  For voltages beyond the collapse point, $V < V_{c,2}$, the total capacitance can be found through minimization of $U$ under the condition $n_2 = 0$, which yields
\be 
C_{dep,2}(V) \simeq \sqrt{\frac{Na^3}{4 \pi (\gamma - V/V_{im})}} \frac{\varepsilon S_2}{a} , \hspace{0.5cm} (V < V_{c,2}).
\ee
This capacitance is similar to Eqs$.$ (\ref{eq:Cdep}) and (\ref{eq:Cdep1}), and is determined by the growing depletion layer at the larger electrode.

\subsubsection{Highly asymmetric capacitor}

We now consider the opposite limit of electrode areas, when $S_2$ and $S_1$ are so different that $S_2/S_1 \gg 1/\sqrt{Na^3}$.  In this case there is no apparent collapse in the capacitance at any negative voltage $-V_{im} < V < 0$ (see the dotted line in Fig$.$ \ref{fig:area_ratios}).  Rather, Eqs$.$ (\ref{eq:VQS}) and (\ref{eq:CQS}) imply a continuous increase in capacitance with voltage as in Eq$.$ (\ref{eq:Cdiv}):
\begin{eqnarray}
C(V) & \simeq & \frac43 \left( \frac{2}{5 \alpha} \right)^{2/3} \left( \frac{V_{im}}{V_{c,1} - V} \right)^{1/3} \frac{\varepsilon S_1}{a}, \nonumber \\ 
& & \hspace{2.5cm} (-V_{im} < V < V_{c,1}). \label{eq:Cdiv2}
\end{eqnarray}
Notice that at $V = -V_{im}$ the capacitance becomes approximately equal to the Helmholtz value $C(-V_{im}) \simeq C_{H,1}$.

At larger negative voltages $V < -V_{im}$ our description of the total energy loses its validity, since Eq$.$ (\ref{eq:VQS}) implies a near-complete filling of this electrode by bound ions, $n_1a^2 \simeq 1$.  Thus the assumption of a dipole-dipole interaction between bound ions is no longer accurate.  At $V \ll -V_{im}$, the capacitance of the smaller electrode should be dominated by the accumulation of complete ionic layers at the small (\emph{negative}) electrode, as described by the theories of Refs$.$ \cite{Freise1952, Borukhov, Nguyen, Kilic2007, Kornyshev2007, Oldham2008}.  This is opposite to the result of the symmetric case, where the capacitance at large negative voltages is determined by the growth of the depletion layer next to the \emph{positive} electrode.  Instead, the large difference between $S_2$ and $S_1$ guarantees that the capacitance of the smaller electrode is much lower, and it therefore determines the total.  The corresponding capacitance at such large negative voltages is
\be 
C(V) \simeq \sqrt{ \frac{- V_{im}}{4 \pi V} } \frac{\varepsilon S_1}{a} , \hspace{0.5cm} (V \ll -V_{im}). \label{eq:Cacc}
\ee
At $V = -V_{im}$, this capacitance also approaches the Helmholtz value $C(-V_{im}) \simeq C_{H,1}$, so that Eqs$.$ (\ref{eq:Cdiv2}) and (\ref{eq:Cacc}) match at $V \simeq -V_{im}$.

Eqs$.$ (\ref{eq:Cdiv2}) and (\ref{eq:Cacc}) indicate that in the limit $S_2/S_1 \gg 1/\sqrt{Na^3}$ the capacitance does not depend on the area of the large electrode $S_2$. Rather, it is totally determined by the physics of the double-layer near the small electrode. Therefore, we can view this result as the capacitance of a single small electrode ($S_2 \rightarrow \infty$). While the result of Eq$.$ (\ref{eq:Cacc}) is known \cite{Freise1952, Borukhov, Nguyen, Kilic2007, Kornyshev2007, Oldham2008}, we could not find that of Eq$.$ (\ref{eq:Cdiv2}) in the literature . In the range of its validity, the capacitance grows with increasing voltage $V$ from $C \sim C_{H,1}$ at  $V = -V_{im}$ and actually diverges at $V = V_{c,1}$. Thus the effective thickness $d^*$ of a single interface vanishes at $V = V_{c,1}$!  Of course, this result is
valid only at zero temperature. As shown in next section, finite
temperature truncates this divergence.

\section{Temperature Dependence of Capacitance} \label{sec:T}

In the previous sections, we derive capacitance by minimization of the total electrostatic energy and do not include any entropic effects associated with the finite thermal energy of ions.  Here we consider the dependence of capacitance on temperature.

Until now, bound ions have been assumed to occupy their ground state configuration: a two-dimensional Wigner crystal.  Since the capacitance is limited only by the resulting dipole-dipole interaction between ions, our prediction for the capacitance is highly dependent on the validity of this description.  If thermal motion of bound ions produces a significant correction to the repulsive interaction between them, than the capacitance will be different than our ``zero-temperature" prediction of the previous sections.

In general, when the two electrodes have area $S_1, S_2$, the total free energy of the system can be written as
\begin{eqnarray}
F & = &  (S_1 n_1 + S_2 n_2) \gamma u_{im} + U_{dep} + U_{dE} + \nonumber \\
& & S_1 f(n_1) + S_2 f(n_2) - QV, \label{eq:F}
\end{eqnarray}
where $f(n)$ represents the free energy per unit area of the two-dimensional collection of bound ions that forms at a given interface, which includes dipole-dipole repulsion and thermal motion along the metal surface.  At low temperatures $k_BT \ll u_{dd}(n_0)$, bound ions are crystallized and $f(n) \simeq \alpha n u_{dd}(n)$, so that Eq$.$ (\ref{eq:F}) becomes equal to Eq$.$ (\ref{eq:Utot2}) and we regain the results of our low-temperature theory.  If the average free energy per ion along each metal surface is small enough that $df/dn \ll u_{im}$, and the voltage is small enough that we still have $n_1, n_2 \ll 1/a^2$, then the equilibrium requirements $\partial F/\partial n_1 = \partial F/\partial n_2 = 0 $ imply
\begin{eqnarray} 
n_0 - n_1 & = & Q/eS_1 \label{eq:n1QT}  \\ 
n_2 - n_0 & = & Q/eS_2, \label{eq:n2QT}
\end{eqnarray}
as before.

As a consequence, for sufficiently small voltages $|V| < V_c$, where neither metal surface is depleted of bound ions, the sum $n_1 + n_2 = 2n_0$ remains constant.  The condition $\partial F/\partial Q = 0$, along with $\partial n_1/\partial Q = -1/eS_1$ and $\partial n_2/\partial Q = -1/eS_2$, gives
\be 
eV = f'(n_2) - f'(n_1). \label{eq:Vf}
\ee
Here, the $\prime$ denotes a derivative with respect to the argument.  The capacitance $dQ/dV$ is therefore
\be 
C = e^2 \left[\frac{f''(n_1)}{S_1} + \frac{f''(n_2)}{S_2} \right]^{-1}.
\label{eq:Cn1n2T}
\ee

Generally speaking, Eq$.$ (\ref{eq:Cn1n2T}) can be combined with Eqs$.$ (\ref{eq:n1QT}) -- (\ref{eq:Vf}) to give an analytic relation for the capacitance as a function of voltage at a given temperature: $C(V, T)$.
This section will focus primarily on the temperature dependence of the zero-voltage capacitance $C(0,T)$ and on the capacitance maximum $C_{max}(T)$.

\subsection{Zero-voltage capacitance}
According to Eq$.$ (\ref{eq:Cn1n2T}), the capacitance at zero voltage can be written
\be 
C(0, T) = \frac{e^2}{f''(n_0)} \left( \frac{1}{S_1} + \frac{1}{S_2} \right)^{-1}, \label{eq:C0f}
\ee
so the capacitance is fully determined by the (temperature-dependent) free energy of ions along the metal surface, $f(n)$.  The behavior of this contribution can be separated into three regimes of temperature.

\subsubsection{Low Temperature}

At very low temperature $k_BT \ll u_{dd}(n_0)$, one can imagine that each ion in the Wigner crystal undergoes small thermal oscillations in the confining potential created by its neighbors.  If this potential is expanded to second-order in the displacement $r$ from the potential energy minimum, then the average squared thermal displacement is 
\be 
\langle r^2 \rangle \approx \frac{\varepsilon k_BT}{13.5 e^2 a^2 n^{5/2}},
\ee 
again assuming a triangular lattice of dipoles.  The positional entropy of a bound ion, relative to an unbound state in the bulk, can be estimated as $\ln (\langle r^2 \rangle/a^2) - \ln (1/Na^3)$.  The free energy per unit area $f(n)$ of bound ions is therefore
\be 
f(n) \simeq n \left[ \alpha u_{dd}(n) - k_BT \ln \left( N \langle r^2 \rangle a \right) \right],
\ee
The corresponding zero-voltage capacitance
\be 
C(0, T) = C(0, 0) \left[ 1 + \frac{2}{3 \alpha} \frac{k_BT}{u_{dd}(n_0)} \right]^{-1}. \label{eq:Clow}
\ee
where $C(0, 0)$ is the zero-voltage capacitance described by Eq$.$ (\ref{eq:C0S}).

\subsubsection{Intermediate Temperature}

At sufficiently large temperatures that $k_BT \gg u_{dd}(n_0)$ but $k_BT \ll u_{im}$, the crystal-like order of dipoles is destroyed, and bound ions are better described as a two-dimensional ideal gas than as a Wigner crystal.  In this limit, the free energy per unit area of bound ions $f(n)$ can be approximated as that of a two-dimensional ideal gas, 
\be 
f_{id}(n) = -n k_BT \ln \left( \frac{N a^3}{n a^2} \right). \label{eq:fid}
\ee
Here we again define the entropy of an ion relative to the bulk.

The description of Eq$.$ (\ref{eq:fid}) assumes that ions are non-interacting, so that their free energy is determined purely by entropic motion.  The effect of relatively weak interaction between ions can be included by a virial expansion of the free energy, 
\be
f \simeq f_{id} + n^2 k_BT B(T).  \label{eq:fint}
\ee
Here, $B(T)$ is the second virial coefficient, calculated from the dipole-dipole interaction energy $u(r)$ between bound ions as
\begin{eqnarray}
B(T) &= & \frac12 \int_0^{\infty} \left(1 - e^{-u(r)/k_BT} \right) 2 \pi r dr, \\
& \approx & 2.65 \left(\frac{e^2 a^2}{\varepsilon k_BT} \right)^{2/3}.
\end{eqnarray}

By Eq$.$ (\ref{eq:C0f}), the resulting capacitance is
\be 
C(0, T) = \frac{e^2 n_0}{k_BT} \left[ 1 + 5.3 \left(\frac{2 u_{dd}(n_0)}{k_BT} \right)^{2/3} \right]^{-1} \left( \frac{1}{S_1} + \frac{1}{S_2} \right)^{-1}. \label{eq:Cint}
\ee 
If the temperature is low enough that $k_BT \ll u_{im} (Na^3)^{1/3}$, then all other corrections to the capacitance beyond that of the virial coefficient are parametrically smaller in $Na^3$.

One can estimate the transition temperature $T_1$ between Eqs$.$ (\ref{eq:Clow}) and (\ref{eq:Cint}) by equating them, which gives
\be 
T_1 \approx 7.9 \alpha^{3/5} u_{dd}(n_0)/k_B.
\ee
As expected, the transition occurs when the thermal energy is of the same order as the dipole-dipole interaction energy.

\subsubsection{High Temperature}

At much larger temperatures $k_BT \gg u_{im}$, ions no longer bind to the metal surface.  Since at these temperatures the change in potential at a given electrode is small compared to the thermal energy, the attraction of ions to the metal produces only a small perturbation in the overall ion density.  Ions therefore form a diffuse screening layer around each metal surface, with a width equal to the Debye-H\"{u}ckel screening radius
\be 
r_s = \sqrt{ \frac{\varepsilon k_BT}{4 \pi e^2 N} }.
\ee

The resulting capacitance per unit area of the double-layer is $\varepsilon/4 \pi r_s$, so that the total capacitance of the two double-layers in series is
\be 
C_{DH} = \frac{ \varepsilon}{4 \pi r_s} \left( \frac{1}{S_1} + \frac{1}{S_2} \right)^{-1}.  \label{eq:CDH}
\ee 
The crossover temperature $T_2$ between the Debye-H\"{u}ckel capacitance $C_{DH}$ and the intermediate-temperature value can be found by equating Eqs$.$ (\ref{eq:Cint}) and (\ref{eq:CDH}), which gives
\be 
T_2 = 2 \gamma u_{im}/k_B.
\ee

\begin{figure}[htb]
\centering
\includegraphics[width=0.45 \textwidth]{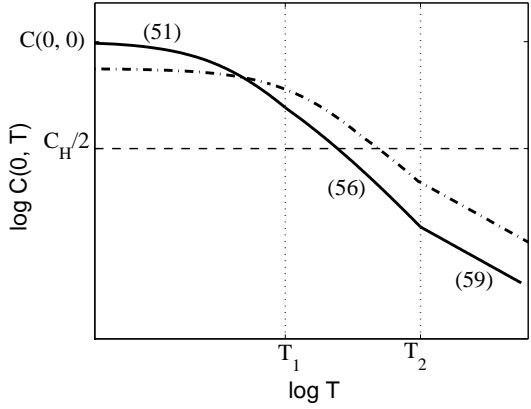}
\caption{Schematic depiction of the temperature dependence of the zero-voltage capacitance of a parallel-plate capacitor.  Numbers in parentheses indicate the formula describing each portion of the temperature dependence.  The dash-dotted line corresponds to a higher value of the density $Na^3$, which produces a lower zero-temperature capacitance and a higher capacitance at large temperatures.  The temperature $T_1$ is indicated for the solid line only.}
\label{fig:C0-T}
\end{figure}

Fig$.$ \ref{fig:C0-T} shows a schematic depiction of $C(0)$ in all three regimes of temperature, plotted for two different values of the density $Na^3$.

\subsection{Maximum capacitance}

At zero temperature, the capacitance maximum is associated with a sharp discontinuity in the $C$-$V$ curve, as shown in Figs$.$ \ref{fig:c-n} and \ref{fig:area_ratios}.  At finite temperature, however, entropic effects will tend to smooth out these discontinuities, thereby shifting both the magnitude and position of the capacitance maximum.

At sufficiently low voltages that neither electrode is depleted of bound ions and $n_1, n_2 \ll 1/a^2$, Eq$.$ (\ref{eq:Cn1n2T}) implies that the capacitance is extremized when
\be 
f'''(n_1) = \left( \frac{S_1}{S_2} \right)^2 f'''(n_2).  
\label{eq:CmaxT}
\ee 
This relation can be used to find the capacitance maximum $C_{max}(T)$ at temperatures $T < T_2$.  For high temperatures $T > T_2$, the capacitance is equal to $C_{DH}$ and is essentially constant in voltage.

Below we derive the maximum capacitance for the symmetric case, $S_1/S_2 = 1$, and the highly asymmetric case, $S_1/S_2 \rightarrow 0$.

\subsubsection{Symmetric capacitor}

When the electrode areas are equal, $S_1 = S_2$, Eq$.$ (\ref{eq:CmaxT}) becomes 
\be 
f'''(n_1) = f'''(n_2).  \label{eq:CmaxTsym}
\ee
This suggests that at $V = 0$, where $n_1 = n_2 = n_0$, there is always either a maximum or a local minimum in the capacitance.  At low temperatures $T \ll T_1$, $V = 0$ is a local minimum.  The maximum can be found by solving Eq$.$ (\ref{eq:CmaxTsym}) and then substituting the results for $n_1$, $n_2$ into Eq$.$ (\ref{eq:Cn1n2T}), which gives
\begin{eqnarray} 
C_{max}(T) & \simeq & C_{max}(0) \left( 1 + 1.2 \frac{ (k_BT/\alpha u_{im})^{1/3} }{\sqrt{n_0 a^2}} \right)^{-1}. \nonumber \\
& & \hspace{35mm} (T \ll T_1)
\label{eq:Cmaxsym}
\end{eqnarray}
Here, $C_{max}(0)$ is the maximum capacitance at zero temperature, given in Eq$.$ (\ref{eq:Cmax}).

Qualitatively, this result can be explained by considering that the capacitance maximum at zero temperature is driven by a vanishing dipole-dipole interaction at the depleted electrode, which allows the capacitance of that electrode to diverge.  At finite temperature, the free energy of bound ions cannot fall below the thermal energy $k_BT$, so the capacitance of the nearly-depleted electrode remains finite.  Setting $u_{dd}(n_1) = k_BT$ and solving for $n_1$, while setting $n_2 = 2 n_0$, allows one to derive the result of Eq$.$ (\ref{eq:Cmaxsym}) to within a numerical coefficient multiplying the temperature.  Note that as $T$ approaches $T_1$, the capacitance maximum approaches $C(0, T)$ as in Eq$.$ (\ref{eq:Clow}).

For larger temperatures $T \gg T_1$, the capacitance maximum disappears and the function changes concavity around $V = 0$.  Thus, the maximum capacitance becomes equal to the zero-voltage capacitance given in Eq$.$ (\ref{eq:Cint}).  A characteristic set of capacitance-voltage curves corresponding to this range of temperature is shown in Fig$.$ \ref{fig:C-V-symmetric}.

\subsubsection{Highly asymmetric capacitor} \label{sec:asymmax}

For the case $S_2 \gg S_1$, our zero-temperature theory in the previous section predicts a sharp divergence in the capacitance as $V$ approaches $V_{c,1}$ [Eq$.$ (\ref{eq:Cdiv})], driven by a vanishing dipole-dipole repulsion at the smaller electrode.  At finite temperature, this divergence is truncated by entropic effects, which inhibit the complete depletion of bound ions from the electrode surface.  According to Eq$.$ (\ref{eq:CmaxT}), for $S_1/S_2 \rightarrow 0$ the maximum is characterized by 
\be 
f'''(n_1) = 0.  \label{eq:CmaxTasym}
\ee
By Eq$.$ (\ref{eq:Cn1n2T}), the corresponding capacitance $C = e^2 S_1/f''(n_1)$.

At low temperatures $T < T_1$, the solution of Eq$.$ (\ref{eq:CmaxTasym}) gives a capacitance 
\be 
C_{max}(T) = \frac{0.32}{\alpha} \left( \frac{u_{im}}{k_BT} \right)^{1/3} \frac{ \varepsilon S_1}{a}, \hspace{5mm} (T \ll T_1).
\label{eq:CmaxT1sym}
\ee 
As in the symmetric case, this maximum occurs when $n_1$ declines sufficiently that $u_{dd}(n_1) \simeq k_BT$.  In the limit that the temperature $T$ approaches $T_1$, $C_{max} \simeq C(0, T)$, and there is no increase in capacitance at positive voltage.

At intermediate temperatures $T_1 < T < T_2$, the capacitance $C(V)$ at small voltages $|V| \ll k_BT/e$ is dominated by the two-dimensional entropy of ions bound to the metal surface $S_1$.  As a consequence, the capacitance increases with negative voltage, where the density of ions $n_1$ increases and therefore their entropy declines.  The capacitance continues to rise with negative voltage until $V \simeq -k_BT/e$, at which point the density of ions is large enough that the dipole-dipole repulsion $u_{dd}(n_1)$ is comparable to the ideal gas free energy per ion $df_{id}(n_1)/dn_1$.  At this point the capacitance achieves its maximum, which is again well-described by Eq$.$ (\ref{eq:CmaxT1sym}).  
Fig$.$ \ref{fig:C-V-asymmetric} shows a characteristic set of $C$-$V$ curves corresponding to this range of temperature.

At large temperatures $T > T_2$, the maximum capacitance becomes similar to the Helmholtz value $C_{H,1}$.  The maximum occurs at large negative voltages $V < -k_BT/e < -V_{im}$, where the applied voltage is strong enough to collapse the ionic screening layer to the electrode surface and form a complete layer.  

\subsection{Comparison with ion-conducting glass experiment}

\begin{figure}[htb]
\centering

\includegraphics[width=0.45 \textwidth]{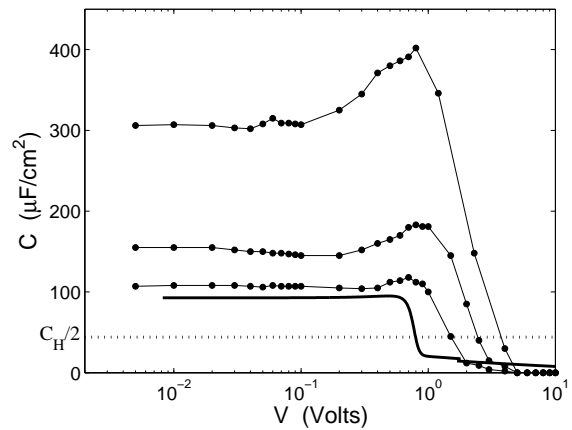}
\caption{The capacitance of an ion-conducting glass between metal
plates.  The thin lines with dots show data from Ref$.$ \cite{Mariappan} from three different samples (reproduced with the authors' permission).  The heavy solid line is our theoretical prediction, using $Na^3 = 0.1$, $a = 2 \Ang$, and $\gamma = 4$.  The dotted line shows the corresponding Helmholtz value $C_H/2$.}
\label{fig:compare_to_glass}
\end{figure}

Let us return to the case of a symmetric double-sided capacitor and compare our theory to the experiments of Ref$.$ \cite{Mariappan}. Capacitance-voltage characteristics for three different phosphosilicate glasses are shown in Fig$.$ \ref{fig:compare_to_glass} together with our theoretical prediction for the relevant concentration $Na^3 = 0.1$ and temperature $T = 600$ K (heavy solid line). The theoretical curve is derived by a numerical minimization of the total free energy, as in Eq$.$ (\ref{eq:F}), using $\varepsilon = 10$ for the bulk of the glass and $\varepsilon = 2.5$ for ions bound to the metal surface.  This approximation is equivalent to using $\gamma = 4$.  If one assumes a uniform dielectric constant $\varepsilon$, or $\gamma = 1$, then the theory predicts an even larger capacitance $C(0) \approx 3 C_H$ (see Fig.\ \ref{fig:CvsT}), but it also predicts the capacitance to collapse at a smaller voltage than what is seen in experiment.  To obtain better agreement with experiment one may need to consider the disorder potential acting on mobile ions in the glass, but this is outside the scope of the present work.

\section{Monte Carlo Simulation}\label{sec:MCSimulation}

In order to test the analytic predictions of the previous sections, we present here the results of a simple Monte Carlo (MC) simulation that models the behavior of an ionic conductor between metal electrodes. As a computational convenience, we divide the system depicted in Fig$.$ \ref{fig:schematic} into three slabs and disregard the thick neutral middle one, so that more computer time may be devoted to the anode and cathode sections in which interesting physics is occurring. This separation is in line with the above theory, which assumes that the two electrodes are separated by a distance much larger than any screening length scale, such that there is no interaction between the two electrodes' double-layers. 

Each remaining slab is treated as a square prism cell with volume $\Omega=L_{x}\times L_{y} \times L_{z}$, where $L_{x}=L_{y}$ and $L_{z}$ is chosen so that the system is at least twice as thick as the depletion layer. The metallic electrode coincides with one of the cell's square faces. At $V=0$, there is no difference between the two partial systems; each contains $\Omega N$ mobile positive ions, which are modeled as spheres with diameter $a = 2$\AA$ $ that carry a charge $e$ located at their center (the ``primitive model"). These mobile ions are constrained to move on a cubic lattice with lattice constant $a$, placed so that $a/2$ is the distance of closest approach of an ion to the cell's walls. Mobile ions are subject to an excluded volume constraint, so that two of them cannot occupy the same site simultaneously. On this same lattice, at each of the $M_{b}=\Omega/a^3$ total lattice sites, there is a small fixed charge $-e Na^3$ which models the negative background. The MC program allows for $Q/e$ ions to be taken from the anodic cell to the cathodic cell without changing the charge of the background. As we will see, this movement of $Q/e$ ions is equivalent to applying a certain positive voltage $V$.

Every charge within a cell forms an electrostatic image in the metallic electrode surface ($z = 0$), \textit{i.e.} a charge $q$ at $(x,y,z)$ has an image charge $-q$ located at $(x,y,-z)$. The total electrostatic energy $\mathcal{E}$ of the cell is calculated as $1/2$ times the energy of a system twice as large composed of the real charges and their images, so that
\begin{equation} \mathcal{E} = \frac{e^{2}}{4 \varepsilon} \sum_{i,j; d_{ij}\neq 0}^{M_{t}}\frac{q_{i} q_{j}}{d_{ij}}.
\label{eq:ElectroE}
\end{equation} 
Here, $q_{i}$ denotes the charge of particle $i$, $d_{ij}$ denotes the distance between particles $i$ and $j$, and $M_t = 2(M_i + M_b)$ is the total number of particles in the system.  For real mobile ions, $q_i=e$; for the mobile ions' images, $q_i=-e$; for the fixed background charges, $q_i=-eNa^3$; and for the images of the background charges, $q_i=eNa^{3}$. The dielectric constant is set to $\varepsilon=5$ everywhere. 

At the beginning of a MC simulation both the temperature $T$ and the zero-voltage mobile ion density $Na^3$ are set. The positive ions are then initialized to random non-overlapping coordinates on the lattice and the initial energy is calculated from Eq$.$ (\ref{eq:ElectroE}). After selecting an ion at random, the MC program attempts to reposition it to a random lattice site within a cubic volume of $(4$ \AA$)^{3}$ centered on the ion's current position. For one in every 100 attempted moves the MC program expands this volume to $(20$ \AA$)^3$ in order to overcome the effects of any large, local energy barriers.  The cell is given periodic boundaries, so that an ion exiting one face of the cell re-enters at the opposite face. The total electrostatic energy of the system, $\mathcal {E}$, is calculated after each attempted move. Moves are then accepted or rejected based on the standard Metropolis algorithm. To ensure thermalization, $8,000$ moves per mobile ion are attempted before any simulation data is collected. After thermalization, simulations attempt between $10^5$ and $10^6$ moves per mobile ion in the system, of which $\sim 10\%-40\% $ are accepted.

\begin{figure}
\includegraphics[width=0.45\textwidth]{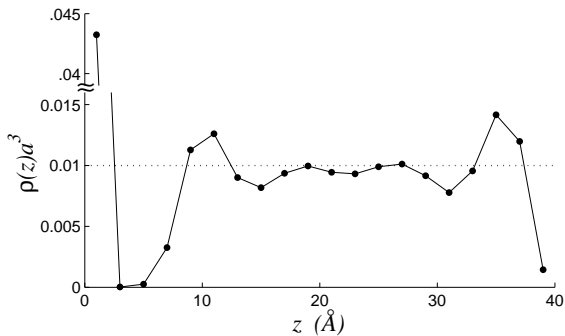}
\caption{The average dimensionless ion density $\rho a^3$ as a function of the distance $z$ from the metal surface, as calculated from a MC simulation of a $40 \times 40 \times 40$ \AA$^3$ cell at $Q = 0$.  The dotted line shows the concentration which neutralizes the negative background.  } \label{fig:density_profile}
\end{figure}

In Fig$.$ \ref{fig:density_profile}, the average ion density as function of the $z$ coordinate is shown for a simulation of a $40 \times 40 \times 40$ \AA$^3$ cell, with $T=350K$ and $Na^3=0.01$. The accumulation of bound ions at the metal surface ($z=1$\AA) is clear, as is the depletion layer adjacent to it. As predicted, the system regains electroneutrality beyond the depletion layer, validating our assumption that the two double layers of the metallic electrodes can be simulated separately.  The peaks in ion density at either edge of the neutral region likely correspond to over-charging by a strongly-correlated liquid of ions in the bulk \cite{Loth, chargeinversion}. 

The primary goal of the simulation is to obtain the full system's capacitance, $C=dQ/dV$. A change in voltage in the real system requires charge transfer, $Q$, from one electrode to the other through the voltage source. In our simulation we apply a constant voltage $V \geq 0$ to the pair of cells by adding $Q/e$ ions to one cell (cathodic side) and removing $Q/e$ ions from the other cell (anodic side). In this way the electronic charge of each electrode is varied; the addition of $Q/e$ ions to a cell corresponds to an addition of a charge $-Q$ to the electrode, which comes in the form of image charges for the added ions. The voltage that corresponds to this movement of charge is found through $V=d \mathcal {F}/dQ$, where $\mathcal {F}= F_{a} + F_{c}$ is the free energy of the full system, equal to the the sum of the anodic and cathodic free energies. In a given cell containing $M_{i}$ positive ions, we used the Widom particle insertion method \cite{Widom1978, Svensson1988} to obtain the change in either the anodic or cathodic cell's free energy $\Delta F_{a,c}$ caused by the addition of another positive ion,
\begin{equation}
\frac{\Delta F_{a,c}}{k_B T}
= \ln\left( \rho(x,y,z) a^{3}/\langle \exp [-\Delta
\mathcal {E}(x,y,z)/(k_B T)]\rangle\right). \label{eq:Widom}
\end{equation}
Here, $\Delta \mathcal {E} (x,y,z)$ is the change in electrostatic energy due to a probe charge placed at $(x,y,z)$ and $\rho(x,y,z)$ is the mean density of real positive ions at $(x,y,z)$. The angle brackets denote a time average and the ratio of the quantities inside the natural log is independent of the position $(x,y,z)$.  At the beginning of the simulation, $20$ lattice sites are selected as Widom insertion sites at which the quantities $\rho(x,y,z)$ and $\exp [-\Delta \mathcal {E}(x,y,z)/(k_B T)]$ are calculated after every attempted move.  The corresponding values of $\Delta F_{a,c}$ obtained from each insertion site are averaged in order to give a final value.

We are free to set the free energy of the system at $Q=0$ to zero, so that $\mathcal {F}(0)=0$. Simulating the anode and cathode for $Q=1,2,3...$, while employing the Widom insertion method, allowed us to determine the system's free energy as a function of $Q$,
\begin{equation}
\mathcal {F}(Q)=\mathcal {F}(Q-1)+\Delta F_c (Q)-\Delta F_a (Q). \label{eq:FreeEnergy}
\end{equation}
This equation must be used iteratively to find $\mathcal {F}(Q-1)$ starting from $\mathcal {F}(0)=0$. For the symmetric case ($S_1=S_2$), this process is easily extended to negative voltages by taking $Q<0$; positive ions are then attracted to the anode and repelled from the cathode. In the asymmetric case ($S1<S2$), $\Delta F_c(Q) = \Delta F_a (-Q S_1/S_2)$, so we need only to simulate the cathodic cell at positive and negative $Q$ in order to calculate $\Delta F_a(Q)$ and $\mathcal {F}(Q)$.
 
Taking the discrete derivative of these data points gives $\mathcal {F}(Q)-\mathcal {F}(Q-e)=V(Q-e/2)$. Another derivative gives the capacitance of the system as a function of $V$, 
\begin{equation}
C (V(Q))=\frac{dQ}{dV}=\frac{e}{(V(Q+e/2)-V(Q-e/2))},
\label{eq:Cap}
\end{equation}
where $V(Q)=[V(Q+e/2)+V(Q-e/2)]/2$.

\begin{figure}
\includegraphics[width=0.5\textwidth]{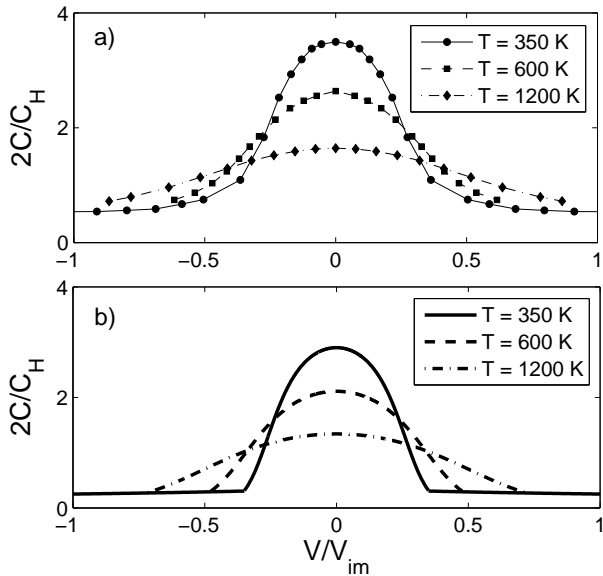}
\caption{Capacitance as a function of voltage for a symmetric parallel-plate capacitor at various temperatures $T_1 \lesssim T < T_2$, using $Na^3 = 0.01$, $\varepsilon = 5$, and $\gamma = 1$.  a) Results from a Monte Carlo simulation of the OCP model.  Error bars for the MC data are smaller than the symbols.  b) Analytic predictions, as explained in section \ref{sec:T}.} \label{fig:C-V-symmetric}
\end{figure}

Fig$.$ \ref{fig:C-V-symmetric} shows the results of the simulation for a system with $S1=S2=40 \times 40$ \AA$^2$, $L_z=20$\AA, and $Na^3=0.01$.  Capacitance as a function of voltage is shown for three values of the temperature, along with the analytic predictions explained in section \ref{sec:T}.  These temperatures fall within the range $T_1 \lesssim T < T_2$, and as predicted, the maximum capacitance occurs at zero voltage while the capacitance collapse is smeared over a voltage range proportional to $k_BT/e$.  

In order to quantify the finite-size effects of our simulation cell, we
examined the capacitance at zero voltage, $C(0, T)$, obtained from three
``slab-shaped" simulation volumes of size $L \times L \times L/2$, with $L =$ 40, 60, and 80 \AA. For $Na^3=0.01$, $C(0, T)$ was seen to scale linearly with $1/L$ at all values of the temperature that we examined ($T = $ 350, 600, and 1200 K).  In each case, the value of $C(0, T)$ obtained by extrapolation to infinite system size was within $16\%$ of the value of $C(0, T)$ corresponding to $L = 40$ \AA.  This difference was within the uncertainty of our simulation for all temperatures except 1200 K.  We also checked that there was no dependence of the capacitance on the aspect ratio of our simulation cell by examining $C(0, T)$ in three cubic cells with side length $L =20$, $40$, and $60\Ang$. The resulting value of $C(0, T)$ again scaled linearly with $1/L$ and, within uncertainty, the extrapolated values of $C(0, T)$ agreed with those found in the slab geometry.  These results allow us to conclude that the $40 \times 40 \times 20$ \AA ~simulation cell provides a good approximation of an infinite system. All MC results presented below correspond to this choice.  

\begin{figure}[htb]
\includegraphics[width=0.5\textwidth]{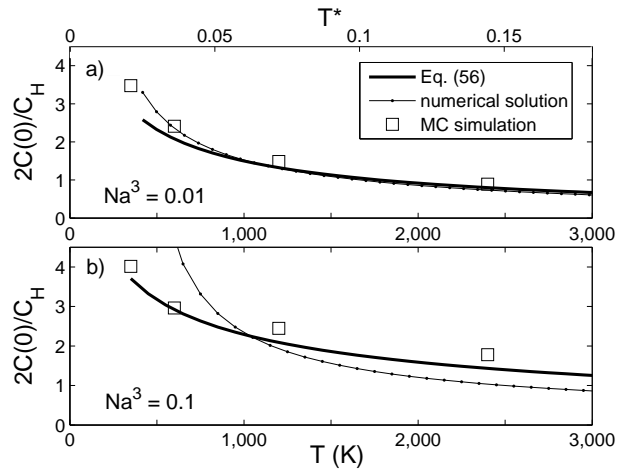}
\caption{Capacitance at zero voltage, $C(0, T)$, as a function of temperature for a symmetric system $S_1 = S_2$ at two different values of the density $Na^3$. Squares represent the MC data using $S=40 \times 40 $\AA$^2$, while the prediction of Eq$.$ (\ref{eq:Cint}) is shown by the thick black curve. Results are plotted as a function of absolute temperature (lower axis) as well as dimensionless temperature $T^* = k_BT/(e^2/\varepsilon a)$ (upper axis).  The error bars for the MC data are smaller than the symbols.
a) $Na^3 = 0.01$. 
b) $Na^3 = 0.1$.}
\label{fig:CvsT}
\end{figure}

The temperatures explored by our MC simulation fall in the intermediate temperature range $T_1 \lesssim T < T_2$, and so the capacitance should be described by Eq$.$ (\ref{eq:Cint}).  Indeed, as shown in Fig$.$ \ref{fig:CvsT}, the temperature dependence of $C(0, T)$ obtained from simulations is in good agreement with the analytical prediction of Eq$.$ (\ref{eq:Cint}).  Results from a numerical minimization of the total free energy [Eq$.$ (\ref{eq:F})] are also shown.  Results are plotted as a function of absolute temperature as well as dimensionless temperature $T^* = k_BT/(e^2/\varepsilon a)$.

Unfortunately, the low-temperature predictions of our theory, corresponding to $T \ll T_1$, could be not be examined directly since these temperatures correspond to an extremely low acceptance rate in our MC simulation.  Nonetheless, we can get an idea of the zero-temperature capacitance by examining the behavior of the total electrostatic energy $\mathcal{E}$.  Since at zero temperature the total free energy becomes equal to $\mathcal{E}$, the capacitance $C$ approaches $(d^2\mathcal{E}/dQ^2)^{-1}$ at low temperatures.  Examining $(d^2\mathcal{E}/dQ^2)^{-1}$ as a function of temperature and extrapolating to $T = 0$ allows us to make a rough estimate of the zero-temperature capacitance $C(0, 0)$.  For $Na^3 = 0.01$, the result is $2C(0,0)/C_H \approx 7$, which is significantly higher than the theoretical prediction of $2C(0,0)/C_H = 4.5$ given by Eq$.$ (\ref{eq:C0}).  This enhanced capacitance may be the result of screening of the dipole interaction by mobile ions in the bulk, which suppresses the interaction of distant dipoles and therefore reduces the effective value of $\alpha$.  At larger ion density, the discrepancy between our low-temperature theory and the projected zero-temperature capacitance from simulation becomes even more pronounced.  For $Na^3 = 0.03, 0.1, 0.3 $ we estimate $2C(0,0)/C_H = 6.5 \pm 1$, as compared to $2C(0,0)/C_H = 3.5, 2.5, 2.0$ given by Eq$.$ (\ref{eq:C0}). These results are consistent with the interpretation based on screening of dipole-dipole repulsion by ions of the bulk. Indeed, at larger ion densities the bulk becomes more effective at screening because it is separated from the metal surface by a thinner depletion layer.

As a rudimentary test of the extent to which the dipole interaction is screened by bulk ions, we performed a MC simulation of a $40 \times 40 \times 20 \Ang^3$ simulation cell in which one of the ions was fixed to the center of the metal surface ($x = y = 0$, $z = a/2$).  The time-averaged density of ions $\rho(x,y,z)$ was then recorded at every lattice site in the simulation cell, from which the mean electric potential $\phi(x,y,z=a/2)$ at the metal surface could be reconstructed.  In order to isolate the contribution of bulk ions to the potential from that of strongly-correlated ions on the metal surface, the simulation was run with $n_0 S$ ions removed from the system.  In this way we simulated the anodic side of a capacitor at $V = V_c$, where the metal surface is depleted of bound ions.  At a density corresponding to $Na^3 = 0.01$ and a temperature $T = 350$ K, we found that the potential $\phi$ surrounding the ion was well described by $\phi(r, z = a/2) = e a^2 / 2 \varepsilon r^3 \cdot \exp(-r/r_{sb})$, where $r = \sqrt{x^2 + y^2}$ is the azimuthal distance from the bound ion and $r_{sb}$ is a length scale which characterizes the range of the screened dipole potential.  We found $r_s \approx 4.5 a$, which is smaller than the average distance $n_0^{-1/2} \approx 6a$ between bound ions at zero voltage.  If this expression $e \phi(n^{-1/2})$ is substituted for the dipole-dipole energy $u_{dd}$, then the resulting prediction for zero-voltage capacitance at $Na^3 = 0.01$ is significantly enhanced: $2C(0)/C_H = 12$ as compared to $4.5$ from Eq.\ (\ref{eq:C0}).  In reality, the observed capacitance from MC simulations is between these two values, $2C(0)/C_H \approx 7$, and this discrepancy may be the result of a non-additive response of bulk charges to dipoles at the surface
(non-linear screening).

\begin{figure}[htb]
\includegraphics[width=0.5\textwidth]{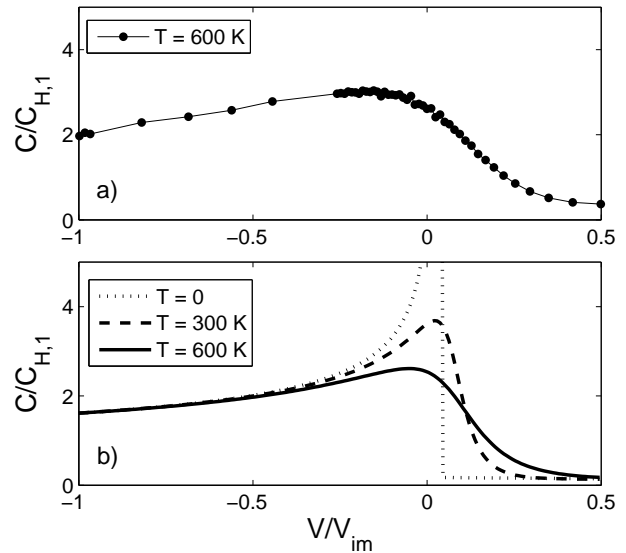}
\caption{The capacitance of a highly asymmetric capacitor, $S_2/S_1 \rightarrow \infty$, where only surface $1$ determines the capacitance.  a) Results from a MC simulation.  Error bars are smaller than the symbols.  b) Analytic predictions for various temperatures, as explained in section \ref{sec:T}.  Note that the capacitance diverges at $T \rightarrow 0$.} \label{fig:C-V-asymmetric}
\end{figure}

Finally, we also considered the simulation of a highly asymmetric capacitor, where $S_2$ is infinite and $S1=40 \times 40 \Ang^2$.  We again examine the case $Na^3=0.01$, $T=600$K, and use $L_z=20 \Ang$.  The results are shown in Fig$.$ \ref{fig:C-V-asymmetric}, along with the analytic predictions of section \ref{sec:T} for different values of the temperature.

\section{Low-voltage capacitance peak in asymmetric ionic liquids}

Up to this point we have discussed the OCP model primarily as it applies to ion-conducting glasses at moderately high temperatures where only the smallest ion (cation) is mobile.  Qualitative agreement of our predictions with experimental data shows that the OCP model is a reasonable zero-order approximation.

In this section we discuss other possible applications of the OCP model, as mentioned in the introduction. One application which immediately comes to mind is to super-ionic crystals, where only the smallest positive ionic species (such as Na$^+$ or Li$^+$) is mobile. In this case there is no reason for a frozen disorder, so that in super-ionic crystals the OCP model should work even better than in ion-conducting glasses.

Perhaps more interesting is the application to ionic liquids, which have recently attracted considerable attention \cite{Kornyshev2007, Madden-il-1, Madden-il-2, Pinilla-il}. In ionic liquids, both positive and negative ions are mobile. In order to get spontaneous polarization near the electrodes at zero voltage, and thereby obtain the low-voltage peak in capacitance predicted by the OCP model, one should consider a strongly asymmetric ionic liquid.

We begin by considering ionic liquids composed of monovalent cations, which we model as rigid spheres with diameter $a$, and much larger monovalent anions, modeled by rigid spheres with diameter $A \gg a$. For example, one may have in mind the ionic liquid made of Na$^{+}$ cations and large non-coordinating anions such as the ``BARF" anion $([B[3,5-(CF_3)_2C_6H_3]_4]^{-})~$ \cite{Krossing}. In such a liquid anions fill most of the space and form a weakly-compressible negative background. The maximum energy of attraction of the anion to its image in the metal surface, $e^2/2\varepsilon A$, is much smaller than the corresponding image attraction $e^2/2\varepsilon a$ for a cation.
Because of their small size, cations easily move between anions. The cations are strongly attracted to the metal plates and therefore rearrange themselves to form the EDL. Thus, we may assume that anions form an analog of the negative background in the OCP model. The maximum density of such a background is approximately $1/A^3$. Thus, we can expect that the capacitance $C(V, T)$ is similar to what we predict in the OPC model, if for $N$ we use $N \sim 1/A^3$, or in other
words $Na^3 \rightarrow (a/A)^3$. For example, if $A/a = 100^{1/3} = 4.6$ we should get a capacitance $C(0, T)$ similar to the case $Na^3 = 0.01$ studied above for the OCP model.

In order to verify these predictions, we ran MC simulations of the
primitive hard sphere model of an ionic liquid between equal-sized metallic electrodes.  In these simulations, monovalent cations and anions are given a diameter $a$ and $A$, respectively, and placed with a particular volume density $N$ in a square prism simulation cell with volume $\Omega = L \times L \times L/2$.  The electrode surface is again chosen to coincide with the $z = 0$ plane, so that an ion of charge $q = \pm e$ whose center is at position $(x,y,z)$ has an image charge $-q$ at $(x,y,-z)$.  The voltage of the electrode is varied by changing the number of cations $M_c$ in the system, as in simulations of the OCP model, while the number of anions $M_a = \Omega N$ remains constant.  The corresponding electronic charge in the electrode is $Q = e(M_a - M_c)$ and the capacitance $dQ/dV$ can be determined from the resulting voltage.  
A very similar simulation method was used by previous authors \cite{Henderson, Bhuiyan} to examine the capacitance of ionic liquids, but the effect of asymmetric ion size was not explored.

Since in this case anions are mobile, unlike in the OCP model, a change in voltage should correspond to a changing number of anions as well as cations in the vicinity of the metal surface.  Therefore, one may object to our method of modifying the charge of the electrode by changing only the number of cations in the system.  For example, one may imagine inducing a charge $Q = -2e$ in the electrode by democratically adding one cation and removing one anion from the simulation cell rather than by adding two cations.  The difference between these two methods, however, is only an infinitesimal change in the bulk ion densities of the simulation cell; the physics of the metal interface is not affected.  To ensure that the capacitance in our simulation is independent of the method of charge transfer, we repeated all of our simulations using both a method where the charge $Q$ is modified by changing only the number of anions $M_a$ and a method where $M_a$ and $M_b$ are changed simultaneously by equal and opposite amounts.  No noticeable change was observed to any of the results presented below.

The microscopic rules of the simulation are identical to those of the OCP model, except that ions are not constrained to move on a lattice and there is no fixed negative background.  The energy of a particular configuration of ions is also identically calculated, with the exception that the hard-core repulsion between ions should be added explicitly to Eq.\ (\ref{eq:ElectroE}).  That is,
\be 
\mathcal{E} = \frac14 \sum_{i,j}^{M_t} u(d_{i,j}),
\ee
where $M_t = 2(M_a + M_c)$ is the total number of charges in the system (ions plus images), and the two-particle interaction energy $u(d_{i,j})$ is
\be 
u(d_{i,j}) = \begin{cases} 
\infty, & d_{i,j} < (D_i + D_j)/2 \\
q_i q_j /\varepsilon d_{ij}, & d_{i,j} > (D_i + D_j)/2
\end{cases}.
\ee
Here, $D_i$ denotes the diameter of ion $i$; $D_i = a$ for cations and $D_i = A$ for anions.  

In addition to the method of Eq.\ (\ref{eq:Cap}) for calculating capacitance, where the voltage is inferred from the change in free energy of the system, for these simulations we used a method where the voltage is measured directly for a given value of $Q$, so that determination of the free energy is unnecessary for calculating capacitance.  The voltage is measured by defining a ``measurement volume" near the back of the simulation cell --- occupying the range $-L/4 < x < L/4$, $-L/4 < y < L/4$, $L/4 < z < 3L/8$ --- inside of which the electric potential is measured.  After performing thermalization of the initial random configuration (50,000 moves per mobile ion), the total electric potential $\phi(x,y,z)$ is measured at 500 equally-spaced points within the measurement volume after every $3(M_a + M_c)$ attempted moves.  These measured values of potential are then averaged both temporally and spatially to produce a value for the electric potential $\bar{\phi}(Q)$ of the electrode relative to the bulk.  The corresponding voltage between the two electrodes is $V = \bar{\phi}(Q) - \bar{\phi}(-Q)$, and the capacitance is determined from the discrete derivative $\Delta Q/\Delta V$.  The results produced by this second method were compared with those produced by the method of the previous section for four different sets of simulation parameters, and the results were indiscriminable.  Below we present results from only the second, more time effective method.

\begin{figure}[htb]
\includegraphics[width=0.5\textwidth]{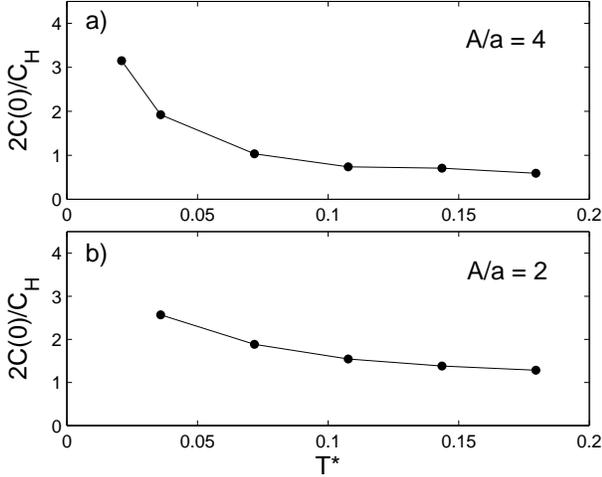}
\caption{The capacitance of a primitive model ionic liquid between metallic electrodes as a function of temperature.  The density of ions in each case is $\rho_b^* = 0.5$.  Error bars are smaller than the symbols. a) $A/a = 4$.  b) $A/a = 2$.  Compare these results to those of the OCP model in Fig.\ \ref{fig:CvsT}}. \label{fig:C-T-ionliquids}.
\end{figure}

Fig.\ \ref{fig:C-T-ionliquids} shows the resulting capacitance for ion liquids with asymmetry $A/a = 4$ and $A/a = 2$ at various temperatures.  Here the temperature is presented in dimensionless units $T^* = k_BT / (e^2/\varepsilon a)$ in order to facilitate comparisons with literature \cite{Bhuiyan, Henderson, Boda}.  Our simulations use the same values for the cation size and dielectric constant as in the OCP case, $a = 2 \Ang$ and $\varepsilon = 5$, so that the temperature scale $e^2/\varepsilon a k_B \approx 16700 \textrm{ K}$ and the range of data $0.02 < T^* < 0.18$ corresponds to $350 \textrm{ K} \leq T \leq 3000 \textrm{ K}$.  For a more typical value of the cation diameter $a \sim 8 \Ang$, this range corresponds to $100 \textrm{ K} \lesssim T \lesssim 750 \textrm{ K}$.  The size of the simulation cell for the $A/a = 4$ case was $L = 80 \Ang$, and in the case $A/a = 2$ we used $L = 40 \Ang$.  The dimensionless ion density $\rho_b^* = M_a (a^3 + A^3)/\Omega$ was $\rho_b^* = 0.5$.  An examination of finite size effects, as in the previous section, suggests that our results for capacitance are accurate to within 18\%.  As predicted above, our numerical results for ionic liquids with $A/a = 4$ and $2$ are close to the results for the OCP model with $Na^3 = 0.01$ and $0.1$, respectively.

\begin{figure}[htb]
\includegraphics[width=0.5\textwidth]{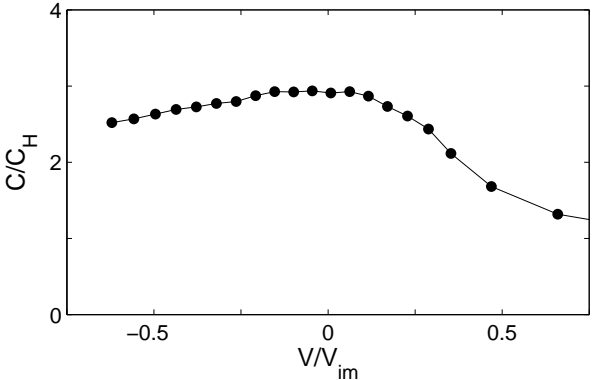}
\caption{The capacitance of a single metal/ionic liquid interface as a function of voltage for an ionic liquid with $A/a = 2$.  The temperature is $T^* = 0.036$ and the density is $\rho_b^* = 0.5$. Error bars are smaller than the symbols.  Compare these results to those of the OCP model in Fig.\ \ref{fig:C-V-asymmetric}.} \label{fig:C-V-il-asymmetric}
\end{figure}

While Fig.\ \ref{fig:C-T-ionliquids} shows the capacitance of the two-electrode system, the capacitance of a single interface can also be easily determined from our MC simulations by looking at the derivative $d\bar{\phi}/dQ$.  As an example, Fig.\ \ref{fig:C-V-il-asymmetric} shows this capacitance as a function of dimensionless voltage $V/V_{im}$ for the case $A/a = 2$ at a temperature $T^* = 0.036$ and density $\rho_b^* = 0.5$.  The asymmetry in capacitance with voltage is similar to what we observed in the OCP model (Fig.\ \ref{fig:C-V-asymmetric}).

The results of Fig.\ \ref{fig:C-V-il-asymmetric} can be compared to those of a recent study \cite{Kornyshev-il} which performed a molecular dynamics simulation of an ionic liquid with size ratio $A/a = 2$ near a uniformly charged interface.  In this study, the capacitance was found to be similarly asymmetric with voltage, with the larger capacitance resulting when the voltage has the same sign as the larger ion.  However, since the study considered only the response of the ionic liquid to a uniformly-charged plane (essentially treating the metal electrode as a charged insulator), it arrived at capacitances $C < C_H$ at all voltages.  It is also worth emphasizing that over the range of voltage in Fig.\ \ref{fig:C-V-il-asymmetric} we are still far from complete coverage of the electrode surface.  Indeed, at $V/V_{im} = 0.5$ the area coverage of the surface is only $35\%$ and $0.2\%$ by anions and cations, respectively, while at $V/V_{im} = -0.5$ it is $3\%$ and $28\%$.  Thus, in our case the collapse in capacitance is not caused by the building of a second layer of ions, as proposed by Ref.\ \cite{Kornyshev2007}.

Thus far we have focused on the effects of asymmetric ion size, but we note that there is another way to make a strongly asymmetric ion liquid.  Namely, cations and anions may have the same radius $a$ but different absolute values of charge. One can imagine, for example, that cations are multivalent and have charge $+Ze$ while anions have charge $-e$. In this case cations are much more strongly attracted to their $-Ze$ images, so that together they again create a dipole layer on the surface of the metal.  Because there are $Z$ anions per one cation, anions form a thicker negatively-charged layer centered farther from the metal surface than the cations.  This anion layer is analogous to the depletion layer of in the OCP theory, with a dimensionless concentration of cations $Na^3 \sim 1/(Z+1)$.  In order to estimate the capacitance, we can use the results of the OCP model with $Na^3 \rightarrow 1/(Z+1)$.

\begin{figure}[htb]
\includegraphics[width=0.5\textwidth]{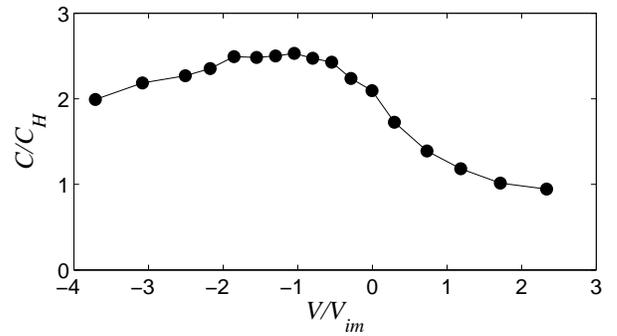}
\caption{The capacitance of a single metal/ionic liquid interface as a function of voltage for an ionic liquid with trivalent cations and monovalent anions with equal diameter.  The dimensionless temperature $T^* = k_BT/(Z^2 e^2/\varepsilon a) = 0.11$ and the density is $\rho_b^* = 0.5$. Error bars are smaller than the symbols.  Compare these results to those of the OCP model in Fig.\ \ref{fig:C-V-asymmetric}.} \label{fig:C-V-Z}
\end{figure}

This prediction can be checked by our MC methods by simulating an ionic liquid with trivalent cations ($Z = 3$) and a neutralizing concentration of monovalent anions, both with the same diameter $a$.  We consider the case where the dimensionless temperature, normalized to the larger charge of the cation, is  $T^* = k_BT/(Z^2 e^2/\varepsilon a) = 0.11$ and the density is $\rho_b^* = 0.5$.  The resulting capacitance is shown in Fig.\ \ref{fig:C-V-Z} as a function of the dimensionless voltage $V/V_{im}$, where $V_{im} = Z e/2\varepsilon a$.  The $C$-$V$ curve is again very similar to that of the OCP model, with a maximum $C/C_H > 1$ and a smaller capacitance at positive voltage, where $Z$-ions are depleted from the electrode surface.  

Even closer imitation of the OCP model can likely be obtained by combining a large charge asymmetry $Z \gg 1$ with a large asymmetry of ion size $A/a \gg 1$.

\section{Aqueous solution of a Z:1 salt}

Another system for which the OCP model gives a zero-order approximation is that of an aqueous solution of a salt with multivalent cations with charge $Ze$ (``$Z$-ions") and monovalent anions, for example LaCl$_3$.  In such a solution $Z$-ions are strongly attracted to their $-Ze$ images in the metal electrode and so a number of them form compact dipoles at the interface. On the other hand, attraction of anions to their images is much weaker, so that at room temperature they do not form dipoles but rather stay in the solution and effectively form a negative OCP model background with charge density $-eZN$, where $N$ is the concentration of salt.  If we again use the theory of Sec.\ \ref{sec:lowT} to balance the depletion layer energy with the energy of Z-ions condensed on the metal surface, we arrive at Eq.\ (\ref{eq:n0}) for the surface concentration $n_0$ of Z-ions. Remarkably, both $n_0$ and the capacitance at $T = V = 0$ do not depend on $Z$ (one can guess this from the fact that the corresponding results in section \ref{sec:lowT} do not depend on the elementary charge $e$).

For LaCl$_3$ we can examine the case where $N = 0.5$ M when the salt is totally dissociated. This concentration corresponds to $Na^3 \sim 0.1$, if for the diameter of the hydrated
La$^{+3}$ ion we use $a = 6 \Ang$. Thus, it is tempting to apply to this case the above finite temperature calculations of capacitance for $Na^3 = 0.1$.

One may worry, however, that the depletion layer, where the
concentration of La$^{+3}$ ions vanishes, is not uniformly charged by Cl$^{-}$ ions with their average density $-3eN$. This may happen because the positive potential $\phi_s$ near the surface of the metal, which develops to balance the image attraction of $Z$-ions to the metal, is much larger than $k_{B}T/e$ and therefore results in the exponential growth of Cl$^{-}$ concentration near the metal surface. For the surface potential we get $Ze\phi_s = Z^{2}e^{2}/2\varepsilon a - k_{B}T\ln(na^2/Na^3)$ , where the second term comes from the entropy that a $Z$-ion loses at the surface in comparison with the bulk solution.  For LaCl$_{3}$, using $a = 6 \Ang$, $\varepsilon= 80$, $T = 300$ K, and $Na^{3} = 0.1$, we get that $Z^{2}e^{2}/2\varepsilon a = 5.3 k_{B}T$ and  $\ln(n_{0}a^2/Na^{3}) < 1$, and therefore, $e\phi_s$ is equal to only $1.7 k_{B}T$. This allows us to ignore (as a zero-order approximation) the non-uniformity of the concentration of Cl$^{-}$ in the depletion layer. Then we can use the temperature-dependent results we obtained for the OCP model in Section \ref{sec:T}. Applying these results requires only the scaling of the temperature in units of $Z^{2}e^{2}/\varepsilon a$. For LaCl$_3$, using $\varepsilon = 80$ and $a = 6 \Ang$, we find that the temperature unit $Z^{2}e^{2}/\varepsilon a$ is 0.37 of that for glass (where $Z=1$, $\varepsilon = 5$, and $a =2 \Ang$). Thus, for $N \sim 0.5$ M at $T = 300$ K we arrive at the same ratio $C(0)/C_H = 2.5$ for the capacitance of a single interface as for the OCP model with $Na^3 = 0.1$ at $T=800$ K (Fig.\ \ref{fig:CvsT}b).

\begin{figure}[htb]
\includegraphics[width=0.5\textwidth]{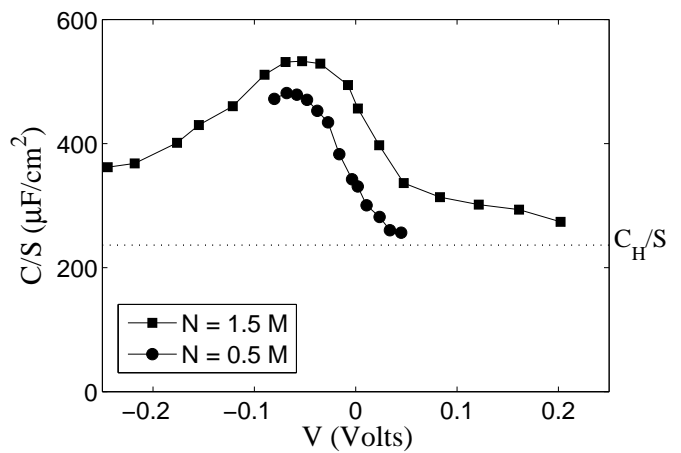}
\caption{The capacitance per unit area $C/S$ of an aqueous solution of LaCl$_3$ at a metal electrode, as determined by our MC simulation using the primitive hard sphere model.  The La$^{+3}$ cation has a hydrated diameter $a = 6 \Ang$, while the Cl$^-$ anion has a diameter $A = 4 \Ang$.  The temperature is $T = 300$ K and the uniform dielectric constant $\varepsilon = 80$.  Error bars are smaller than the symbol size.  The dotted line shows the Helmholtz value $C_H/S \approx 236$ $\mu$F/cm$^2$.  Compare these results to those of the OCP model in Fig.\ \ref{fig:C-V-asymmetric}.} \label{fig:LaCl3}
\end{figure}

This prediction can be checked by running a MC simulation identical to the one described in the previous section.  Fig.\ \ref{fig:LaCl3} shows the resulting capacitance per unit area $C/S$ of a single interface as a function of voltage for the salt concentrations $N = 0.5$ M and $N = 1.5$ M, using the temperature $T = 300$ K and the estimated hydrated diameters $a = 6 \Ang$ and $A = 4 \Ang$ for La$^{+3}$ and Cl$^-$, respectively.  The simulation cell is given a size $L = 100 \Ang$.  For $N = 0.5$ M and $1.5$ M, the maximum capacitance is larger than the Helmholtz value by $2.0$ and $2.3$ times, respectively.  The data presented in Fig.\ \ref{fig:LaCl3} corresponds to the range of electrode charge $|\Delta Q| < Z e N \Omega/2$, where $Z e N \Omega$ represents the total cation charge in the simulation volume.  Restricting our simulation to this range ensures that the bulk ion concentration is not changed significantly by the addition/removal of Z-ions to the cell that is associated with finite voltage.

The results of Fig.\ \ref{fig:LaCl3} are qualitatively similar to those of the OCP model given in Fig.\ \ref{fig:C-V-asymmetric}.  The capacitance is asymmetric with respect to voltage, acquiring a larger value at negative voltages where Z-ions accumulate at the metal surface and bind strongly to their image charges.  At positive voltages, Z-ions are depleted from the electrode surface and the capacitance collapses.  At large positive voltages $V \gg k_BT/e \approx 26$ mV, negative anions become strongly bound to the electrode by the applied voltage and they approach complete filling of an ionic layer.  In this limit the capacitance approaches $\varepsilon / 2\pi A = C_H \cdot a/A$.

\section{Conclusions}

In this paper we have presented a theory to explain how the capacitance of the metal/ionic conductor interface can be significantly larger than the Helmholtz capacitance $C_H$.  In other words, we have shown how the apparent thickness of the double layer $d^*$ can be smaller than the ion radius.  This surprising conclusion is obtained by abandoning the mean-field approximation and considering instead the behavior of discrete charges next to the metal surface.  While mean field theories cannot explain how $d^*$ can be smaller than the physical separation between the electrode and its countercharge, we have shown that very large capacitance is a natural result for an EDL composed of discrete, correlated ions.  We have worked within the approximation of a ``one-component plasma" model, where only one species of ion is mobile, and described its behavior over the full range of temperature.  We have further argued that our results can be easily extended to strongly asymmetric ionic liquids.  A simple Monte Carlo simulation confirms our analytical predictions at realistic temperatures.  At very low temperatures, the EDL capacitance is limited only by the weak repulsion between ion-image dipoles at the metal surface, which for a single interface produces a sharp capacitance peak that diverges as $T^{-1/3}$.  

Qualitatively, our theory explains all the main features of the experiment in Ref.\ \cite{Mariappan}. The authors of Ref.\ \cite{Mariappan} relate their observations to theories of so-called ``pseudo-capacitance", a term used by Conway and coworkers for the rare cases of anomalously large EDL capacitance (see \cite{Conway} and references therein).  Pseudo-capacitance is said to result from specific adsorption of cations to the metal surface, where the cations are neutralized. In this sense Conway's theory is similar to ours.  However, his theory of pseudo-capacitance does not explain what happens with the negative charge of excess anions, which remains in the bulk and which in our theory plays a pivotal role.  The existing theory of pseudo-capacitance also does not  explicitly specify the form of the repulsion between bound ions, and therefore does not arrive at a closed result.  We take care to address both of these points in the present (OCP) model, and we arrive at definite predictions for capacitance.  Thus, one may consider our theory to be an improved theory of pseudo-capacitance, if by this term one understands a capacitance larger than the Helmholtz value. We emphasize, however, that our theory does not assume any Faradaic effects, so that our result is in fact a standard capacitance and the prefix ``pseudo-" is unnecessary.

\begin{small}
\vspace*{2ex} \par \noindent
{\em Acknowledgments.}

We are grateful to M. M. Fogler, T. T. Nguyen, S. D. Baranovskii, M. Palassini, B. Roling, M. Bowring, and C. Varma for helpful discussions.  B. S. acknowledges the support of the NSF. M. S. L. thanks the FTPI for financial support.  B. I. S. is grateful for the hospitality of the Aspen Center for Physics, where part of this work was done.
\end{small}


\end{document}